\documentclass[reprint,%
superscriptaddress,%
twocolumn,%
showpacs,preprintnumbers,
 amsmath,amssymb,
 aps,
]{revtex4}

\usepackage{graphicx}% Include figure files
\usepackage{dcolumn}% Align table columns on decimal point
\usepackage{bm}% bold math
\usepackage{hyperref}% add hypertext capabilities
\begin{document}

\title{Improved dynamics and gravitational collapse of tachyon field \\
 coupled with a barotropic fluid}

\author{Jo\~ao Marto}
\email{jmarto@ubi.pt}
\affiliation{Departamento de F\'{i}sica, Universidade da Beira Interior, 6200
Covilh\~a, Portugal}%
\affiliation{Centro de Matem\'atica e Aplica\c{c}\~oes da Universidade da Beira
Interior (CMA-UBI)}%

\author{Yaser Tavakoli}
\email{tavakoli@ubi.pt}
\affiliation{Departamento de F\'{i}sica, Universidade da Beira Interior, 6200
Covilh\~a, Portugal}%
\affiliation{Centro de Matem\'atica e Aplica\c{c}\~oes da Universidade da Beira
Interior (CMA-UBI)}%
\affiliation{Departamento de F\'{\i}sica, Universidade Federal do Esp\'{\i}rito Santo, Av. Fernando Ferrari 514, Vit\'{o}ria - ES, Brazil}%

\author{Paulo Vargas Moniz}
\email{pmoniz@ubi.pt}
\affiliation{Departamento de F\'{i}sica, Universidade da Beira Interior, 6200
Covilh\~a, Portugal}
\affiliation{Centro de Matem\'atica e Aplica\c{c}\~oes da Universidade da Beira
Interior (CMA-UBI)}%

\begin{abstract}

We consider a spherically symmetric gravitational collapse of a
tachyon field with an inverse square potential, which is coupled with
a barotropic fluid. By employing an holonomy correction imported
from loop quantum cosmology, we analyse the dynamics of the collapse
within a semiclassical description. 
Using a dynamical system approach,
we find that the stable fixed points given by  the  standard general relativistic setting turn into 
saddle points in the present context. %corresponding to the stable ones given  by the standard general relativistic setting. 
This provides a new dynamics in contrast
to  the black hole and naked singularities solutions appearing
in the classical model. 
Our results
suggest that classical singularities can be avoided
by quantum gravity  effects and are replaced by a bounce. 
By a thorough numerical studies  we show that, depending on the barotropic parameter $\gamma$, there exists a 
class of solutions corresponding to either a  fluid or a  tachyon dominated regimes. 
Furthermore, for the case $\gamma\sim1$, we find an interesting tracking behaviour between the tachyon and the fluid leading to a dust-like collapse.
In addition, we show that, there exists a threshold
scale which determines when an outward energy flux emerges, as a non-singular
black hole is forming, at the corresponding collapse final stages.

\end{abstract}


\date{\today}

\maketitle

%\keywords{Loop quantum cosmology; holonomy correction; gravitational collapse.}

%\ccode{PACS numbers: 04.20.Dw, 04.60.Pp, 04.60.Bc}

%% 04.20.Dw,04.60.Bc

%\tableofcontents

\section{Introduction}

The spherically symmetric gravitational collapse, with a variety of
matter fields, has been well studied in general relativity (see Ref.~\cite{Joshi:2007}
and references therein). Those investigations indicate that the gravitational
collapse, depending on the initial conditions, may produce a black
hole with a singularity inside or a naked singularity as its final
state \citep{Joshi:2007,Penrose:1965,Hawking:1974}. However, these
results are not expected to hold in a quantum theory of gravity. Among
the candidates for a theory of quantum gravity, loop quantum gravity
(LQG) \citep{Ashtekar:2004,LQC-AS,Bojowald:2010,Ashtekar:2005} is a non-perturbative
background independent theory. From the effective constraints approach
used in the LQG program, there are two general types of quantum corrections,
namely the `inverse triad' and `holonomy' types.

The study of the gravitational collapse 
has thus been considered in LQG, by means of these corrections.
It was proposed that inverse triad modifications resolve the classical
singularities that arise at the final state of the gravitational collapse,
whose matter source is a standard scalar field \citep{Bojowald:2005,Goswami:2006}.
Moreover, in a homogeneous and spherically symmetric model, loop gravity
effects, within a holonomy correction, modify the standard Friedmann
equation by adding a $-\rho^{2}/\rho_{{\rm crit}}$ correction term
into it. In a cosmological context, these effects resolve the big
bang singularity and replace it by a bounce \cite{Ashtekar:2006}.
In addition, for gravitational collapse of a scalar field \cite{us:2013c},
with non-interacting particles (dust) and a perfect fluid describing
radiation \cite{Modesto:2013}, it was shown that those quantum gravity
effects provide a threshold scale for a non-singular black hole formation.

More recently, the gravitational collapse of a self
interacting tachyon field coupled with a barotropic fluid has been
considered in a classical context \cite{us:2013a}, exploring the
non-canonical features of the tachyon  kinetic term and its subsequent
(anti)friction effects. Therein, a dynamical system analysis was employed,
where, by making use of the specific kinematical features of the tachyon
field (which are rather different from a standard scalar field), it
was established the initial conditions for which either the tachyon
or the fluid becomes dominant. It was found the conditions under which
a black hole or a naked singularity scenarios are produced, as well
as solutions with a tracking behaviour between the tachyon and the
fluid.

A tachyon scalar field has also been investigated in
loop quantum cosmology (LQC) as a concrete example for investigating
the initial singularity of the universe \cite{Li:2009}. In the context
of a  gravitational collapse, by employing quantum gravity effects
of the inverse triad type, it was proposed that the geometry of space-time
near the classical singularity is regular \cite{us:2013b}. Furthermore,
some novel features such as evaporation of the horizons in the presence
of the inverse triad modifications were studied. Nevertheless, holonomy
type correction can bring, in the context of homogeneous LQC, 
distinct and interesting physical aspects when compared with the inverse triad type. 
Therefore, it is of interest to investigate how a modification provided by holonomy
corrections to the tachyon equations of motion can avoid the classical
singularity that may arise at the final state of the gravitational collapse.
In addition, another question that can be addressed here is how this
type of loop quantum effect can indeed affect the emergence of trapped
surfaces in this  kind of models. These questions constitute our main
goal to  be explored in this paper.

The organization of this paper is as follows. In
section \ref{collapse-1} we employ the loop gravity effects with
holonomy corrections to the gravitational collapse of a tachyon field
in addition with a barotropic fluid. Then, in section \ref{collapse-dynamics}
we present a phase space analysis for our semiclassical model and
will find the possible fixed point solutions to the evolution equations. 
This analysis provides a correspondence between the
fixed point solutions found in the herein semiclassical regime and those given by
their general relativistic counterpart presented in Ref.~\cite{us:2013a}.
Nevertheless, it will be shown that the corresponding loop gravity
modified fixed point solutions, due to the holonomy effects, are free
of the central singularity whereas their classical counterpart are not.
In section \ref{collapse-end}, by using numerical techniques, we
will study the evolution of trapped surfaces in the semiclassical
interior space-time of the collapse. This analysis will further provide a contrasting
between our herein semiclassical analysis and the one given by the
inverse triad modified collapsing scenario  presented in Ref.~\cite{us:2013b}.
Finally, in section \ref{conclusion} we present the conclusion and
discussion of our work.

\section{Gravitational collapse: Improved dynamics}

\label{collapse-1}

We consider a spherically symmetric gravitational collapse whose interior
space-time is the marginally bound case, i.e., the $k=0$ FLRW \cite{us:2013a}.
Let $t$ be the proper time for a falling observer whose geodesic
trajectories are labeled by the comoving radial coordinate $r$, and
$R(t,r):=ra(t)$ is the area radius of the collapsing cloud. Then,
for a continuous collapsing scenario, we take $\dot{R}=r\dot{a}<0$ (with a `dot' denoting a derivative with respect to the proper time $t$),
implying that the area radius of the collapsing shell, for a constant
$r$, decreases monotonically.

The corresponding Hamiltonian constraint for the interior geometry
is provided as \cite{Ashtekar:2005} 
\begin{equation}
{\cal H}\ =\ -\frac{3}{\kappa\tilde{\gamma}^{2}}c^{2}\sqrt{|p|}+{\cal H}_{{\rm matt}}\ ,
\label{Hamil-gr}
\end{equation}
where $c:=\tilde{\gamma}\dot{a}<0$ and $p:=a^{2}$ are, respectively,
the conjugate connection and triad satisfying the non-vanishing Poisson
bracket $\{c,p\}=\kappa\tilde{\gamma}/3$, with $\tilde{\gamma}\approx0.23$
being the Barbero-Immirzi dimensionless parameter. Moreover, $\kappa=8\pi G$,
and ${\cal H}_{{\rm matt}}=\rho V$ is the matter Hamiltonian with
$V$ being the volume of the fiducial cell \cite{Ashtekar:2005}.

A pertinent scenario to investigate semiclassically effects suggested
from LQG (as far as a gravitational collapse is concerned) is the
so-called \emph{holonomy correction}. The algebra generated by the
holonomy of phase space variables $c$ is just the algebra of the
almost periodic function of $c$, i.e., $e^{i\bar{\mu}c/2}$ where
$\bar{\mu}$ is inferred as kinematical length of the square loop,
since its dimension is similar to that of a length, which together
with $p$, constitutes the fundamental canonical variables in quantum
theory \cite{Ashtekar:2005}. This consists semiclassically in replacing
$c$ in Eq.~(\ref{Hamil-gr}), with the phase space function, by
means of 
\begin{align}
\frac{1}{2i\bar{\mu}}\left(e^{i\bar{\mu}c}-e^{-i\bar{\mu}c}\right)=\frac{\sin(\bar{\mu}c)}{\bar{\mu}}~.\label{HolonomyCorrection1}
\end{align}
It is expected that the classical theory is recovered for small $\bar{\mu}$;
we therefore obtain the effective semiclassical Hamiltonian \citep{Ashtekar:2006,Taveras:2006}
\begin{align}
{\cal H}_{{\rm eff}}\ =\ -\frac{3}{\kappa\tilde{\gamma}^{2}\bar{\mu}^{2}}\sqrt{|p|}\sin^{2}(\bar{\mu}c)+{\cal H}_{\mathrm{matt}}\ .\label{EFFham-1}
\end{align}
The dynamics of the fundamental variables is then obtained by solving
the system of Hamilton equations; i.e., 
\begin{align}
\dot{p} & \ =\ \{p,{\cal H}_{{\rm eff}}\}=-\frac{\kappa\tilde{\gamma}}{3}\frac{\partial{\cal H}_{{\rm eff}}}{\partial c}\notag\\
 & \ =\ \frac{2a}{\tilde{\gamma}\bar{\mu}}\sin(\bar{\mu}c)\cos(\bar{\mu}c).\label{HAMeqs-1}
\end{align}
Furthermore, the vanishing Hamiltonian constraint (\ref{EFFham-1})
implies that 
\begin{equation}
\sin^{2}(\bar{\mu}c)\ =\ \frac{\kappa\tilde{\gamma}^{2}\bar{\mu}^{2}}{3a}{\cal H}_{{\rm matt}}\ .\label{Hamilton-Eq2}
\end{equation}
Thus, using Eqs. (\ref{HAMeqs-1}) and (\ref{Hamilton-Eq2}), we subsequently
obtain the modified Friedmann equation, $H=\dot{a}/a=\dot{p}/2p$~:
\begin{equation}
H^{2}\ =\ \frac{\kappa}{3}\rho\left(1-\frac{\rho}{\rho_{\text{crit}}}\right),
\label{Friedmann-eff-1a}
\end{equation}
where $\rho_{{\rm crit}}=3/(\kappa\tilde{\gamma}^{2}\lambda^{2})\approx0.41\rho_{{\rm Pl}}$,
and $\rho$ is the total (classical) energy density of the collapse
matter content. Eq. (\ref{Friedmann-eff-1a}) implies that the classical
energy density $\rho$ is limited to the interval $\rho_{0}<\rho<\rho_{{\rm crit}}$
having an upper bound at $\rho_{{\rm crit}}$, where $\rho_{0}\ll\rho_{{\rm crit}}$
is the energy density of the star at the initial configuration, $t=0$.
Hence, the effective energy density reads 
\begin{equation}
\rho_{{\rm eff}}\ :=\ \rho\left(1-\frac{\rho}{\rho_{{\rm crit}}}\right).\label{newFriedmann}
\end{equation}
We see that the effective scenario, provided by holonomy corrections,
leads to a $-\rho^{2}$ modification of the energy density, which
becomes important when the energy density becomes comparable to $\rho_{{\rm crit}}$.
In the limit $\rho\rightarrow\rho_{{\rm crit}}$, the Hubble rate
vanishes; a classical singularity is thus replaced by a bounce.

From Eq.~(\ref{Friedmann-eff-1a}) the time derivative of the  Hubble rate reads
\begin{align}
\dot{H}\ =\ -\frac{\kappa}{2}(\rho+p)\left(1-2\frac{\rho}{\rho_{{\rm crit}}}\right), 
\label{LFriedmann2}
\end{align} 
Then, using the relation $\ddot{a}/a=\dot{H}+H^2$ we obtain the modified Raychaudhuri equation as
\begin{align}
\frac{\ddot{a}}{a} \ =\  -\frac{\kappa}{6}\: \rho\left(1-4\frac{\rho}{\rho_{{\rm crit}}}\right)
-\frac{\kappa}{2}\: p\left(1-2\frac{\rho}{\rho_{{\rm crit}}}\right).
\label{Ray-1}
\end{align} 
By redefinition of  the Raychaudhuri equation (\ref{Ray-1}), similar to the corresponding classical relation, in an effective form  $\ddot{a}/a=-\kappa/6(\rho_{\rm eff}+3p_{\rm eff})$,
we obtain the effective pressure $p_{\rm eff}$ of the system  as \citep{Ashtekar:2006,Taveras:2006}
\begin{equation}
p_{\text{eff}}\ =\ p\left(1-2\frac{\rho}{\rho_{{\rm crit}}}\right)-\frac{\rho^{2}}{\rho_{{\rm crit}}}\ .\label{Peff}
\end{equation}
The effective energy density $\rho_{{\rm eff}}$ and pressure $p_{{\rm eff}}$ satisfy the conservation equation: 
%by the relation
\begin{align} 
\dot{\rho}_{\text{eff}}+3H(\rho_{\text{eff}}+p_{\text{eff}}) = 0\ .
\label{cons-eff}
\end{align}
We can  define the effective equation of state as 
\begin{equation}
w_{{\rm eff}}\ :=\ \frac{p_{{\rm eff}}}{\rho_{{\rm eff}}}\ =\ \frac{p}{\rho}\left(\frac{\rho_{{\rm crit}}-2\rho}{\rho_{{\rm crit}}-\rho}\right)-\frac{\rho}{\rho_{{\rm crit}}-\rho}\ .\label{Eq-State-eff}
\end{equation}

In general relativity, the equation for an apparent horizon in a spherical
symmetric space-time is given by $g^{ij}R_{,i}R_{,j}=0$, which corresponds to $F/R=1$;
the $F(R)=\kappa\rho R^{3}/3$ is the mass function of the collapsing
matter. Thus, the space-time is said to be trapped or untrapped if
$F>R$ or $F<R$, respectively \cite{Joshi:2007}.  Since
in the considered effective scenario, the energy density $\rho$
in the Friedmann equation (\ref{Friedmann-eff-1a}) is modified as $\rho_{{\rm eff}}$,
hence, the mass function $F(R)$ should be modified in the herein semiclassical regime and can be written 
in the following form \cite{us:2013c}: 
\begin{equation}
F_{{\rm eff}}\ =\ \frac{\kappa}{3}\rho_{{\rm eff}}R^{3}\ =\ F\left(1-\frac{\rho}{\rho_{{\rm crit}}}\right).\label{massF-eff}
\end{equation}
We followed a possible perspective in effective scenario in which,  the phase space
trajectories are considered to have classical form whereas the matter components take effective form 
due to  quantum effects. The $\rho/\rho_{{\rm crit}}$
term in Eq.~(\ref{massF-eff}) can be written as 
\begin{equation}
\frac{\rho}{\rho_{{\rm crit}}}\ =\ \frac{a_{{\rm min}}^{3}}{a^{3}}\frac{F}{F_{{\rm crit}}}\ ,\label{massF-eff-2}
\end{equation}
where $a_{{\rm min}}$ and $F_{{\rm crit}}:=(\kappa/3)\rho_{{\rm crit}}r^{3}a_{{\rm min}}$
are respectively, values of the scale factor and mass function
at the bounce. It is seen from Eq.~(\ref{massF-eff-2}) that, the
mass function $F$ changes in the interval $F_{0}\leq F\leq F_{{\rm crit}}$
along with the collapse dynamical evolution, so that, it remains finite
during the semiclassical regime; $F_{0}=(\kappa/3)\rho_{0}r^{3}a_{0}^{3}$
is the initial value  for the mass function at $t=0$. Furthermore,
the effective mass function (\ref{massF-eff}) vanishes at the bounce.

We should notice that the effective mass function (\ref{massF-eff}), likewise the effective energy density and pressure, must be consistent with  the continuity equation (\ref{cons-eff}). 
By working out the relations  (\ref{massF-eff}), (\ref{newFriedmann})  and (\ref{Peff}),  using  the conservation equation (\ref{cons-eff}),  
we  find the following effective equations describing the dynamics of our semiclassical gravitational collapse:
\begin{align}
\kappa\rho_{\rm eff}= \frac{F_{{\rm eff},r}}{R^2R_{,r}}\ ,   \quad  \kappa p_{\rm eff}= -\frac{\dot{F}_{{\rm eff}}}{R^2\dot{R}}\ ,   \quad \dot{R}^2=\frac{F_{\rm eff}}{R}\ ,
\label{grav-eff}
\end{align}
with `$,r$' denoting  the derivative with respect to $r$. In  the classical limit 
where $\rho/\rho_{\rm crit}\ll 1$, we have that $\rho_{\rm eff}\rightarrow\rho$;~ $p_{\rm eff}\rightarrow p$ and $F_{\rm eff}\rightarrow F$, 
then,  Eq. (\ref{grav-eff}) reduces to the known  Einsein's equations for   gravitational collapse \cite{Joshi:2007}.
%In LQC, 

Let us follow Refs.~\citep{us:2013a,us:2013b} and consider the total
energy density, $\rho$, of the collapse to be 
\begin{equation}
\rho\ =\ \rho_{\phi}+\rho_{\gamma}\ ,\label{ener-1}
\end{equation}
which constitutes the classical energy densities of the tachyon field
and the barotropic fluid. In a strictly classical setting, the energy
density $\rho_{\phi}$ and pressure $p_{\phi}$ of the tachyon field
are given by 
\begin{equation}
\rho_{\phi}=\frac{V(\phi)}{\sqrt{1-\dot{\phi}^{2}}},~~~~p_{\phi}=-V(\phi)\sqrt{1-\dot{\phi}^{2}}\ ,\label{energyTach}
\end{equation}
%with a dot denoting a derivative with respect to the proper time $t$,
where  $V(\phi)$ is the tachyon potential. Furthermore, the energy density
of the barotropic fluid, $\rho_{\gamma}$, reads 
\begin{equation}
\rho_{\gamma}\ =\ \rho_{\gamma0}\left(\frac{a}{a_{0}}\right)^{-3\gamma},\label{ener-2}
\end{equation}
where $\rho_{\gamma0}$ is a positive constant denoting the fluid
density at the initial configuration, $a_{0}$, of the collapse, and
$\gamma$ is an adiabatic index satisfying $p_{\gamma}=(\gamma-1)\rho_{\gamma}$,
with $p_{\gamma}$ being the pressure of the barotropic fluid.

For a physically reasonable matter content for the collapsing cloud,
the tachyon field and the barotropic fluid would have to satisfy the
weak and dominant energy conditions. It is straightforward to show
that the tachyon matter satisfies the weak and dominant energy conditions.
For a fluid with the barotropic parameter $\gamma>0$,
the weak energy condition is satisfied, however, concerning the dominant
energy condition, it follows that $\gamma$ must hold the range $\gamma\leq2$
\cite{us:2013a}.

From the total energy conservation equation for collapse matter source,
we could write, generally, that 
\begin{align}
 & \dot{\rho}_{\phi}+3H(1+w_{\phi})\rho_{\phi}=-\Gamma_{{\rm int}}\ ,\label{LConserv2}\\
 & \dot{\rho}_{\gamma}+3\gamma H\rho_{\gamma}=+\Gamma_{{\rm int}}\ ,\label{LConserv2-b}
\end{align}
where the function $\Gamma_{{\rm int}}$ denotes the  interaction  between the tachyon and fluid.
%%
%\textcolor{blue}{
A natural interpretation of  $\Gamma_{{\rm int}}$ is that it implies  an energy transfer between the tachyon field and the barotropic fluid.
Such cases were  studied in (classical) cosmological scenarios where  two fluid system  drive an accelerating universe \cite{TB-Int1,TB-Int2,TB-Int3}. % \cite{TB-Int1,TB-Int2,TB-Int3,TB-Int4,TB-Int5}.
Our main goal in this paper is to investigate  the effects of  LQG  holonomy corrections to the tachyon equations of motion,
and consequently,  the  emergence of trapped surfaces in the herein semiclassical collapse.
Furthermore, we will discuss the situation in which only a tachyon matter  or a fluid is present.
Thus, our approach will be to consider the non-interacting case to fully access the underlying dynamics. 
However,  it is expected that for the interacting case the main features of the bouncing scenario will remain unchanged. 
But, allowing a situation where the tachyon field and the barotropic fluid no longer obey a local  conservation equation, 
might certainly change the outcome of the horizon formation in the presence of the interaction term. % case. 
When we find  the solutions for the non-interacting case, we may get a general idea of 
how the presence of the interaction term can affect our  outcome,
however, a full independent  analysis will be required to describe the dynamics of the collapse in the case of  interacting tachyon matter.  
In the present context, we  thus only assume  that the tachyon field is  self interacting,
i.e., $\Gamma_{{\rm int}}=0$.
The conservation energy density (\ref{LConserv2}) for the tachyon
field gives 
\begin{equation}
\ddot{\phi}=-\left(1-\dot{\phi}^{2}\right)\left[3H\dot{\phi}+\frac{V_{,\phi}}{V}\right],\label{field-eq-1}
\end{equation}
where `$,\phi$' denotes the derivative with respect to $\phi$. Furthermore,
the equation of state $w_{\phi}$ for tachyon field is given by 
\begin{equation}
w_{\phi}\ :=\ \frac{p_{\phi}}{\rho_{\phi}}=-\left(1-\dot{\phi}^{2}\right).\label{Eq-State-Tach}
\end{equation}
In addition, one can define a barotropic index for the tachyon fluid:
$\gamma_{\phi}:=(\rho_{\phi}+p_{\phi})/\rho_{\phi}=\dot{\phi}^{2}$.

\section{Holonomy effects and phase space analysis}

\label{collapse-dynamics}

The use of dynamical system techniques to analyse a tachyon field
in gravitational collapse has been considered in Refs. \citep{us:2013b,us:2013a}.
In what follows, a dynamical system analysis of the tachyon field
gravitational collapse within the improved dynamics approach of LQG
will be studied.

We assume the time variable (instead of the proper time $t$ present
in the comoving coordinate system $\{t,r,\theta,\varphi\}$) 
\begin{equation}
N\ :=\ -\ln\left(\frac{a}{a_{0}}\right)^{3},\label{Nvar}
\end{equation}
defined in Ref. \cite{us:2013a}; therein $0<N<\infty$ where the limit
$N\rightarrow0$ corresponds to the initial condition of the collapsing
system ($a\rightarrow a_{0}$), and the limit $N\rightarrow\infty$
corresponds to $a=0$, i.e., the classical singularity identified
in Ref. \cite{us:2013a}. For an arbitrary function $f$ we get 
\begin{equation}
\frac{df}{dN}\ =\ -\frac{\dot{f}}{3H}\ .
\end{equation}
To analyze the dynamical behaviour of the collapse, we further introduce
the following variables: 
\begin{align}
 & x:=\dot{\phi}\ \ \ \ \ \ \ \ \ \ y:=\frac{\kappa V}{3H^{2}}\ ,\ \ \ \ \ \ \ \ z:=\frac{\rho}{\rho_{{\rm crit}}}\ ,\notag\\
 & s:=\frac{\kappa\rho_{\gamma}}{3H^{2}}\ ,\ \ \ \ \lambda:=-\frac{V_{,\phi}}{\sqrt{\kappa}V^{\frac{3}{2}}}\ ,\ \ \ \ \Gamma:=\frac{VV_{,\phi\phi}}{(V_{,\phi})^{2}}\ .\label{DynV2}
\end{align}

The Friedmann constraint (\ref{Friedmann-eff-1a}), in terms of the
new variables (\ref{DynV2}), can be rewritten as 
\begin{equation}
1=\left(\frac{y}{\sqrt{1-x^{2}}}+s\right)(1-z)\ ,\label{DynFrid}
\end{equation}
in which, the dynamical variables $x$, $y$ and $z$ must satisfy
the constraints $-1\leq x\leq1$, $y\geq0$ and $0\leq z\leq1$. Furthermore,
the time derivative of the Hubble rate, Eq.~(\ref{LFriedmann2}),
in terms of dynamical variables (\ref{DynFrid}), can be written 
\begin{equation}
\frac{\dot{H}}{3H^{2}}\ =\ -\frac{1}{2}(1-2z)\left[\frac{x^{2}y}{\sqrt{1-x^{2}}}+\gamma s\right].\label{H-der}
\end{equation}

Using the Eq. (\ref{DynV2}) and the constraint (\ref{DynFrid}),
the equations of state (\ref{Eq-State-Tach}) and (\ref{Eq-State-eff}),
in terms of new variables can be written as 
\begin{align}
w_{\phi}\ = & -\left(1-x^{2}\right)\ ,\\
w_{\text{eff}}\ = & -(1-x^{2})\left(\frac{1-2z}{1-z}\right)\label{DynState}\\
 & +s(\gamma-x^{2})(1-2z)-\frac{z}{1-z}\ .\nonumber 
\end{align}
Moreover, the fractional densities of the two fluids are respectively
defined as: 
\begin{equation}
\Omega_{\phi}:=\frac{\kappa\rho_{\phi}}{3H^{2}}=\frac{y}{\sqrt{1-x^{2}}}\ ,\ \ \ \ \Omega_{\gamma}:=\frac{\kappa\rho_{\gamma}}{3H^{2}}=s\ .
\end{equation}

An autonomous system of equations, in terms of the dynamical variables
of Eq. (\ref{DynV2}), together with Eqs. (\ref{DynFrid}) and (\ref{H-der}),
is then retrieved: 
\begin{align}
\frac{dx}{dN} & =\left(1-x^{2}\right)\left(x-\frac{\lambda}{\sqrt{3}}\sqrt{y}\right),\label{Dynx}\\
\frac{dy}{dN} & =\frac{\lambda x}{\sqrt{3}}y^{\frac{3}{2}}-y(1-2z)\left[\frac{x^{2}}{1-z}+s\left(\gamma-x^{2}\right)\right],\label{Dyny}\\
\frac{dz}{dN} & =z\left[x^{2}+s(1-z)\left(\gamma-x^{2}\right)\right],\label{Dynz}\\
\frac{ds}{dN} & =s\left[\gamma-\left(1-2z\right)\left(\frac{x^{2}}{1-z}+s\left(\gamma-x^{2}\right)\right)\right],\label{Dyns}\\
\frac{d\lambda}{dN} & =\frac{1}{\sqrt{3}}\sqrt{\kappa}\lambda^{2}x\sqrt{y}\left(\Gamma-\frac{3}{2}\right).
\end{align}
Notice that, in the limit $\rho\ll\rho_{{\rm crit}}$ (i.e., in the
absence of $z$), the Eqs.~(\ref{Dynx})-(\ref{Dyns}) reduce to
the corresponding classical autonomous system of equations in Ref.~\cite{us:2013a}.

We will assume the tachyon potential to be of an inverse square form
\citep{us:2013a,us:2013b}: 
\begin{equation}
V(\phi)\ =\ V_{0}\phi^{-2}.\label{pot-tach}
\end{equation}
For the choice (\ref{pot-tach}) we get $\lambda=\pm2/\sqrt{V_{0}}$
and $\Gamma=3/2$, i.e., as constants. The dynamical system will be
four differential equations with variables $(x,y,z,s)$. Let $f_{1}:=dx/dN$,
$f_{2}:=dy/dN$, $f_{3}:=dz/dN$ and $f_{4}:=ds/dN$. Then, the critical
points $q_{{\rm c}}=(x_{c},y_{c},z_{c},s_{c})$ are obtained by setting
the condition $(f_{1},f_{2},f_{3},f_{4})|_{q_{{\rm c}}}=0$. Next
we will study the stability of our dynamical system at each critical
point by using a standard linearization and stability analysis.

To determine the stability of critical points, we need to perform
linear perturbations around each point by using the form $q(t)=q_{{\rm c}}+\delta q(t)$;
this results in the equations of motion $\delta q'={\cal M}\delta q$,
where ${\cal M}$ is the Jacobi matrix of each critical point whose
components are ${\cal M}_{ij}=(\partial f_{i}/\partial q_{j})|_{q_{{\rm c}}}$.
A critical point is called stable (unstable) whenever the eigenvalues
$\zeta_{i}$ of ${\cal M}$ are such that $\mathsf{Re}(\zeta_{i})<0$
($\mathsf{Re}(\zeta_{i})>0$). If neither of the these cases are achieved,
the critical point is called a saddle point \cite{NSDyn}. We have
summarized the fixed points for the autonomous system (\ref{Dynx})-(\ref{Dyns})
and their stability properties in Table \ref{fixed-points}.

\paragraph*{Point $P_{1}$--}

The eigenvalues of this fixed point are $\zeta_{1}=-2$, $\zeta_{2}=-1$,
$\zeta_{3}=+1$ and $\zeta_{4}=\gamma-1$. All characteristic values
of this point are real, but at least one is positive and two are negative,
thus, the trajectories approach this point on a surface and diverge
along a curve; this is a \emph{saddle} point.

\paragraph*{Point $P_{2}$--}

For this fixed point, the characteristic values are $\zeta_{1}=-2$,
$\zeta_{2}=-1$, $\zeta_{3}=+1$ and $\zeta_{4}=\gamma-1$, which
are the same eigenvalues as the fixed point $P_{1}$, and thus, similar
to the $P_{1}$, this is a \emph{saddle} point.

\paragraph*{Point $P_{3}$--}

This fixed point has eigenvalues $\zeta_{1}=0$, $\zeta_{2}=y_{0}^{2}+\lambda^{2}y_{0}/6>0$,
$\zeta_{3}=\gamma_{1}$ and $\zeta_{4}=(\gamma-\gamma_{1})$, where,
$y_{0}:=-\lambda^{2}/6+\sqrt{(\lambda^{2}/6)^{2}+1}$ and $\gamma_{1}:=\lambda^{2}y_{0}/3$.
For $\gamma>\gamma_{1}$ this point is not stable; for $\gamma<\gamma_{1}$
this is a saddle point.

\paragraph*{Point $P_{4}$--}

The eigenvalues read $\zeta_{1}=+1$, $\zeta_{2}=-\gamma$, $\zeta_{3}=+\gamma$
and $\zeta_{4}=-\gamma$. For $\gamma\neq0$, this point possesses
eigenvalues with opposite signs; therefore, this point is saddle.
For the case $\gamma=0$, this point has one real and positive eigenvalue,
and others are zero, so $P_{4}$ is not a stable point.

\paragraph*{Point $P_{5}$--}

This point is located at $(\sqrt{\gamma},3\gamma/\lambda^{2},s_{0})$,
where $s_{0}:=\left(1-\frac{3\gamma}{\lambda^{2}\sqrt{1-\gamma}}\right)$.
The eigenvalues for this fixed point are $\zeta_{1}=0$, $\zeta_{3}=\gamma$
and 
\begin{align}
\zeta_{2,4}=\frac{2-\gamma\pm\sqrt{\left(1-\gamma\right)\left(4-16s_{0}\gamma\right)+\gamma^{2}}}{4}\ .\label{eigencle-eff}
\end{align}
For $\gamma<\gamma_{1}$, all eigenvalues are non-negative, and for
$\gamma=\gamma_{1}$, we have $\zeta_{2}>0$ and $\zeta_{4}=0$. Therefore,
this point is not a stable fixed point. Notice that, since $0<s_{0}<1$,
the barotropic parameter $\gamma$ must hold the range $\gamma\leq\gamma_{1}<1$
for this fixed point.

\paragraph*{Point $P_{6}$--}

The eigenvalues of this point are the same as the point $P_{5}$,
so that this is not a stable point. Furthermore, the existency condition
for this point implies that the barotropic index satisfies $\gamma\leq\gamma_{1}<1$.

\paragraph*{Point $P_{7}$--}

The eigenvalues for this fixed point are $\zeta_{1}=-2$, $\zeta_{2}=-\gamma$,
$\zeta_{3}=\gamma$ and $\zeta_{4}=1-\gamma$. At least one characteristic
value is negative and one is positive, so $P_{7}$ is a saddle point.

\paragraph*{Point $P_{8}$--}

For this point, the eigenvalues are similar to those of point $P_{7}$,
i.e., $\zeta_{1}=-2$, $\zeta_{2}=-\gamma$, $\zeta_{3}=\gamma$ and
$\zeta_{4}=1-\gamma$. Therefore, this is a saddle point.

%%%%%%%%%%%%%%%%%%%%
%\par 
\begin{widetext}
\begin{center} 
\begin{table}[h!]
%\tbl{Summary of critical points and their properties.}{
%{\tiny{}}%
\begin{tabular}{cccccccc}
\hline 
\hline 
Point & $x$ & $y$  & $z$ & $s$ & $\Omega_{\phi}$  & Existence & Stability \tabularnewline
\hline 
$P_{1}$  & $1$  & 0  & $0$  & $0$  & $1$  & all $\lambda$, $\gamma$  & Saddle  \tabularnewline
$P_{2}$ & $-1$  & 0  & $0$  & $0$  & $1$  & all $\lambda$,  $\gamma$  & Saddle \tabularnewline
$P_{3}$  & $\frac{\lambda}{\sqrt{3}}\sqrt{y_{0}}$  & $y_{0}$  & $0$  & 0  & $1$  &~ all $\lambda$, $\gamma$ ~ & Unstable  for $\gamma\ge\gamma_{1}$ \tabularnewline
 &  &  &  &  &  &  & Saddle  for $\gamma<\gamma_{1}$ \tabularnewline
 $P_{4}$  & $0$  & $0$  & $0$  & $1$  & 0  & all $\lambda$, $\gamma$  & Saddle for $\gamma\neq0$ \tabularnewline
 &  &  &  &  &  &  & Unstable  for $\gamma=0$ \tabularnewline
 $P_{5}$ & ~$\sqrt{\gamma}$~ & ~$\frac{3\gamma}{\lambda^{2}}$ ~&~ $0$~ & ~$s_{0}$~  &~ $1-s_{0}$~ & ~all $\lambda$;~ $\gamma<\gamma_{1}<1$ ~ & Unstable \tabularnewline
$P_{6}$  & $-\sqrt{\gamma}$  & $\frac{3\gamma}{\lambda^{2}}$ & $0$  & $s_{0}$  & $1-s_{0}$  & ~all $\lambda$;~ $\gamma<\gamma_{1}<1$  ~& Unstable \tabularnewline
$P_{7}$  & $1$  &  0  & $0$  & $1$  &  $0$   & all $\lambda$, $\gamma$  & Saddle   \tabularnewline

$P_{8}$  & $-1$  & {\tiny{0 }} & $0$  & $1$  & $0$  & all $\lambda$, $\gamma$  & Saddle  \tabularnewline
\hline 
\hline 
% &  &  &  &  &  &  & \tabularnewline
\end{tabular}
\caption{Summary of critical points and their properties }
\label{fixed-points}
%\caption{Summary of critical points and their properties }
\end{table}
\end{center}
\end{widetext}%%%%%%%%%%%%%%%%%

In the standard general relativistic collapse of a tachyon field with
barotropic fluid \cite{us:2013a}, the fixed points $(x_{c},y_{c},s_{c})=(1,0,0)$
and $(x_{c},y_{c},s_{c})=(-1,0,0)$ are stable fixed points (attractors)
and correspond to a tachyon dominated solution; furthermore,
the collapse matter content behaves, asymptotically, as a homogeneous
dust-like matter which leads to a black hole formation at late times
\cite{us:2013a}. Nevertheless,  in the semiclassical
regime presented here, the existence  of the loop
(quantum) holonomy correction term $z\neq0$, turn
these fixed points into saddles, so that the stable points (i.e.,
the singular black hole solution) of the classical collapse disappear.

The points $(x_{c},y_{c},s_{c})=(1,0,1)$ and $(x_{c},y_{c},s_{c})=(-1,0,1)$,
in the classical regime (in the absence of the $z$ term), correspond
to the stable fixed points (attractors), namely the fluid dominated
solutions, and lead to the black hole formation as the collapse end
state \cite{us:2013a}. Nevertheless, holonomy effects, in the presence
of $z$ term induce respectively, the corresponding saddle points
$(x_{c},y_{c},s_{c},z_{c})=(1,0,1,0)$ and $(x_{c},y_{c},s_{c},z_{c})=(-1,0,1,0)$
for the collapsing system, instead. This means that the classical
singular black holes are absent in  this semiclassical
regime.

\begin{figure}[t]
\centering\includegraphics[height=2.5in]{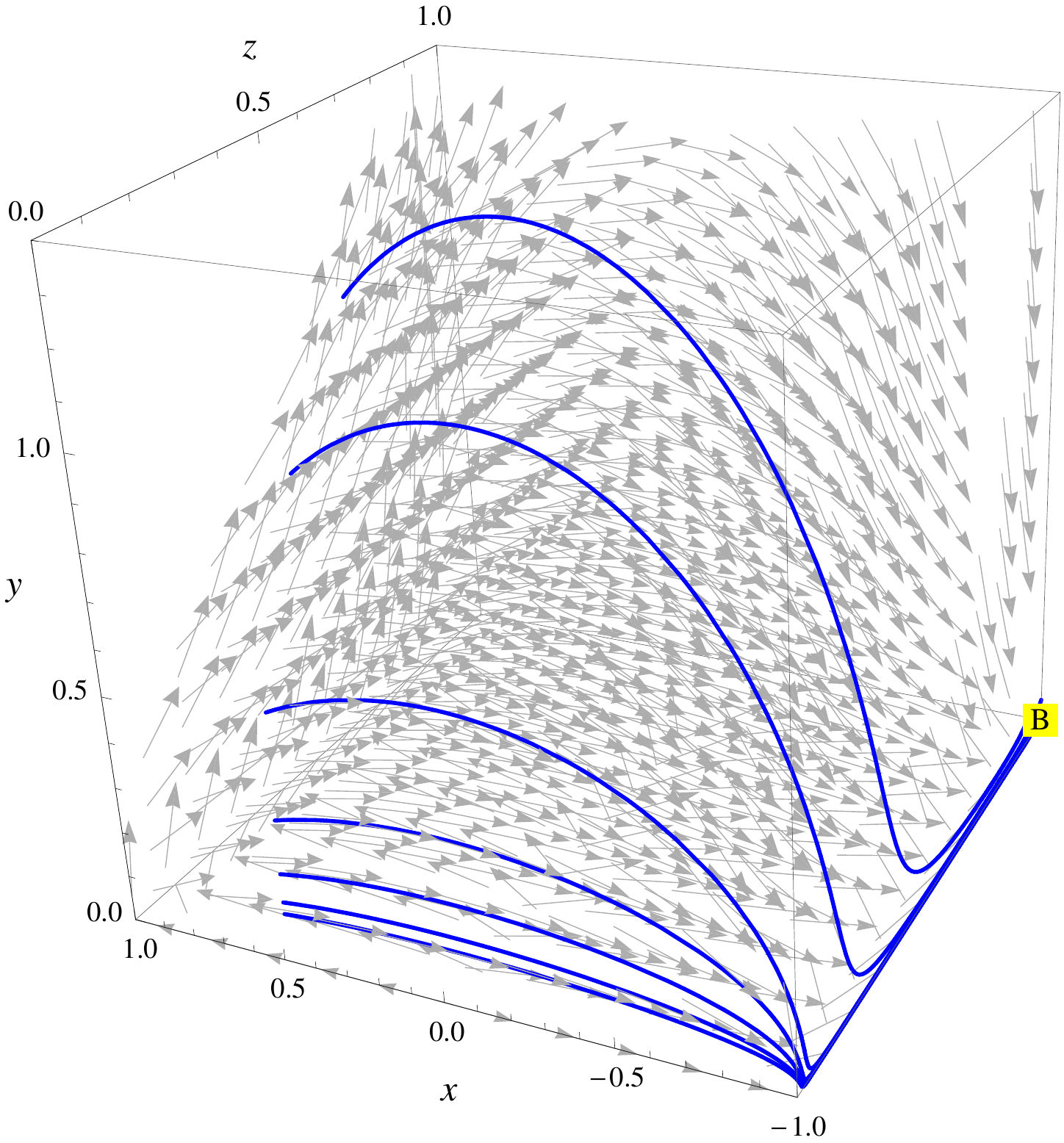} \caption{{\footnotesize{{This plot represents a set of trajectories evolving
in the three dimensional phase space $\left(x,\: y,\: z\right)\equiv\left(\dot{\phi},\:\kappa V/3H^{2},\:\rho/\rho_{\textrm{crit}}\right)$;
also the complete vector field generated by the dynamical system is
shown. All the possible trajectories are tangent to this vector field.
The initial conditions for solving the dynamical system Eqs. (\ref{Dynx})-(\ref{Dyns})
are chosen such that the trajectories start from locations near the
$x-y$ plane. In this plane, $z\approx0$, consequently $\rho\ll\rho_{\textrm{crit}}$.
The different curves are obtained varying the initial value of the
tachyon field $\phi_{0}$. Point B correspond to the location in the
phase space where the semiclassical bounce is defined.}}}}

\label{F-stream3D} 
\end{figure}

In figure \ref{F-stream3D} we show a selection of numerical solutions
of the dynamical system Eqs. (\ref{Dynx})--(\ref{Dyns}), in terms
of the variables $\left(x,\: y,\: z,\: s\right)$. This figure represents
trajectories which start from the lower $x-y$ plane and evolve in
the phase space. These trajectories will initially converge to a point
where $\dot{\phi}\rightarrow-1$, along the $x-y$ plane; however,
in the vicinity of this point, they diverge along the $y-z$ plane
and move away from it. This point can be identified to be the saddle
fixed fixed points $P_{2}$ or $P_{8}$.

However, it is pertinent to point the following. Figure \ref{F-stream3D}
involves parametric functions $x\left(N\right)$, $y\left(N\right)$
and $z\left(N\right)$. The numerical solution shows that the variable
$N$ is only defined on a finite interval $\left[0,\: N_{{\rm bounce}}\right]$;
this can be seen from Eq. (\ref{Nvar}) in which the scale factor
is bounded from below, i.e., $a_{{\rm min}}<a<a_{0}$. In fact, and
contrasting with the classical solution \cite{us:2013a}, where $x\left(N\rightarrow\infty\right)\rightarrow\pm1$
and $y\left(N\rightarrow\infty\right)\rightarrow0$ are asymptotic
limits,   in the semiclassical scenario, the variable
$N$ is bounded at the bounce. This boundary is shown in figure \ref{F-stream3D}
where the curves end at a region where $z\rightarrow1$ (identified
as point B in the plot), which consequently, cannot be classified
as a fixed point of the dynamical system. The numerical study supports
the analytical discussion that the solutions in Ref. \cite{us:2013a} for
points $P_{1}$, $P_{2}$, $P_{7}$ and $P_{8}$, are now avoided
on the semiclassical trajectories. In addition, for all the trajectories
on the phase space shown in figure \ref{F-stream3D}, point $B$ corresponds
to a bouncing scenario which we will analyse it on the next section.

\section{Semiclassical collapse end state}

\label{collapse-end}

In this section we present additional results related to the numerical
studies of Eqs. (\ref{LFriedmann2})--(\ref{field-eq-1}).

\subsection{Tracking solutions: Tachyon versus barotropic fluid}

\begin{figure}
\centering\includegraphics[height=2.2in]{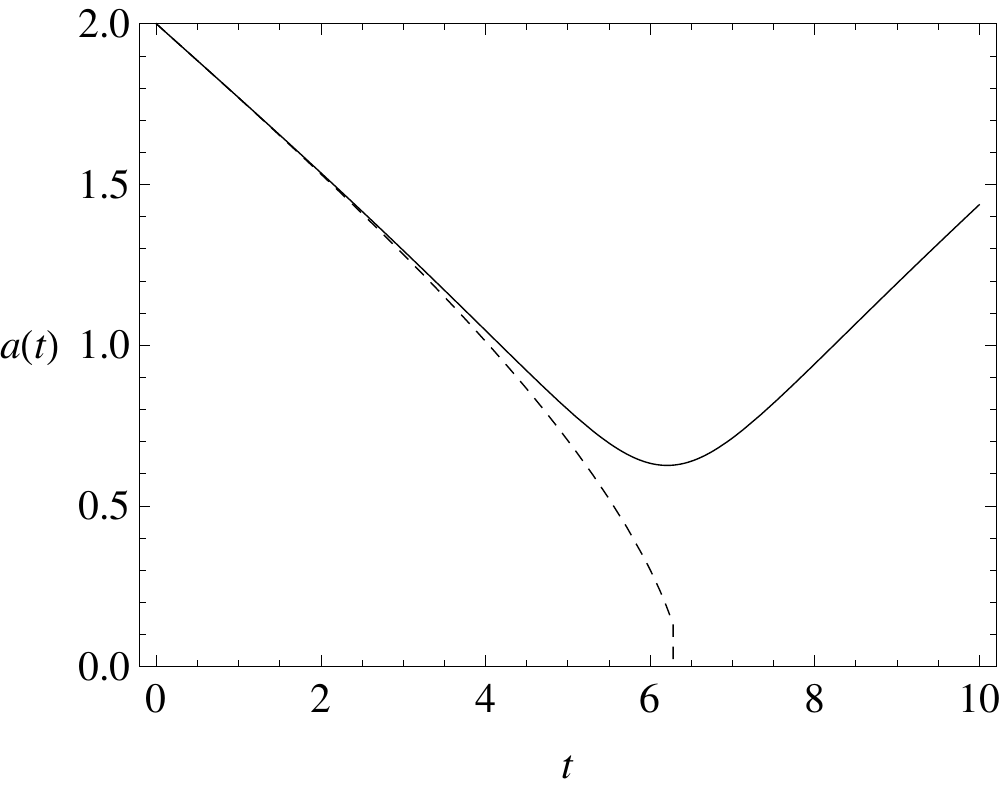}\caption{{\footnotesize{Behaviours of the scale factor $a(t)$ in the semiclassical
(full line), and classical (dashed curve) regime. We considered the
initial conditions as: $\rho_{0}=0.04$, $V_{0}=0.001$, $a(0)=2$,
$\dot{\phi}\left(0\right)=0.5$ and $\phi_{0}=0.6$. We also have
$\gamma=0.1$.}}}

\label{F-scalef} 
\end{figure}

\begin{figure*}[t]
\centering\includegraphics[height=1.9in]{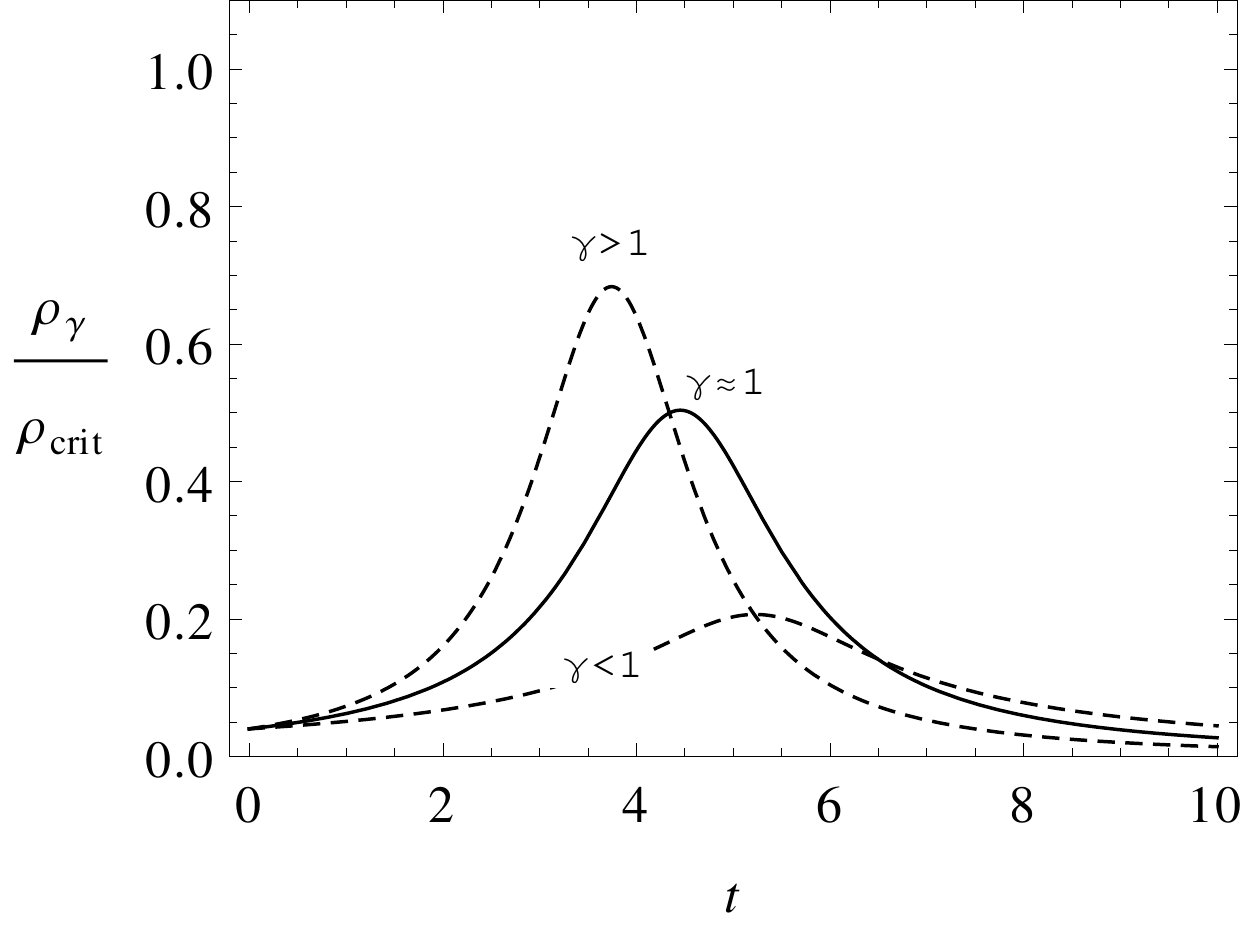}\enskip{} \quad \quad \includegraphics[height=1.9in]{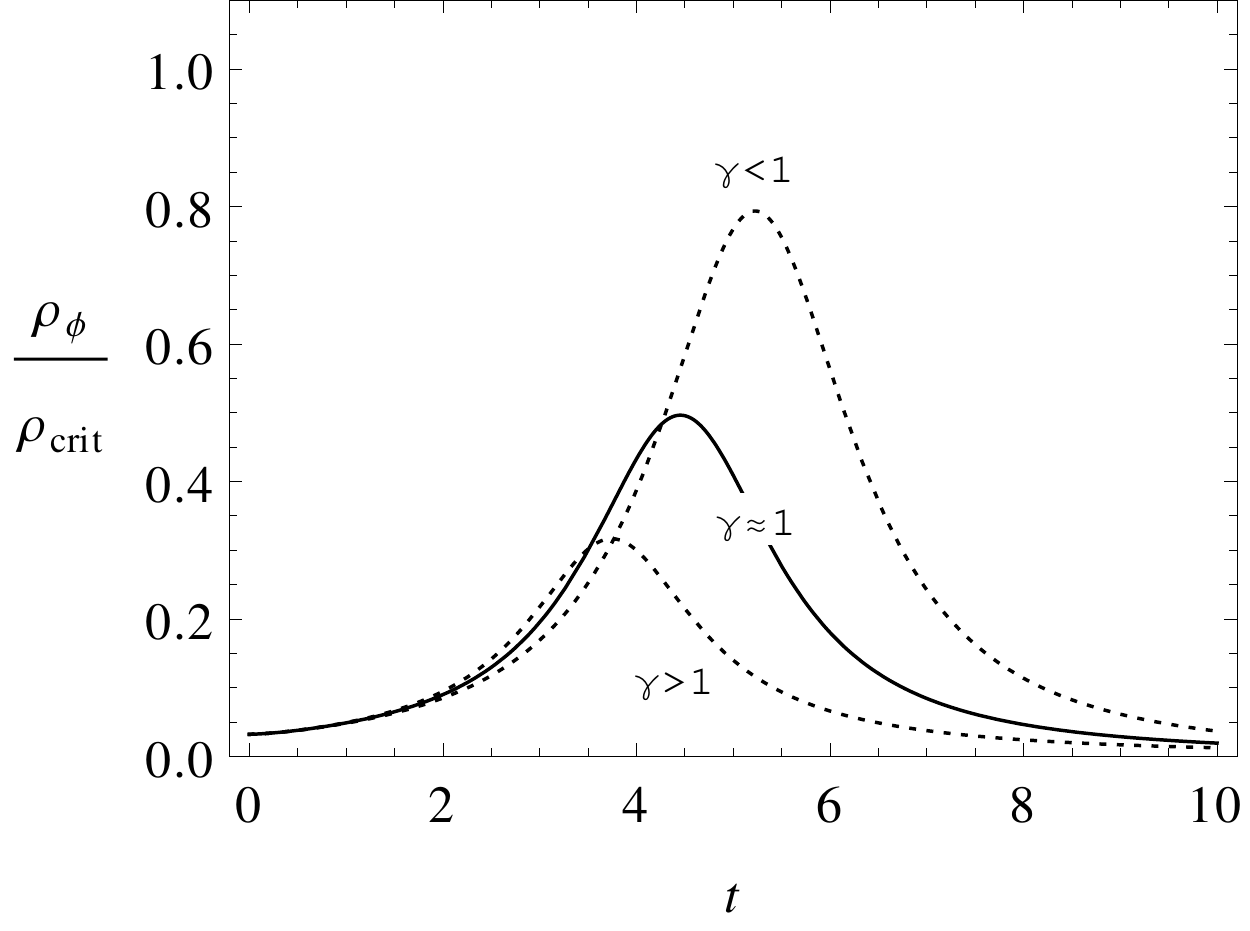} %\enskip{}\includegraphics[height=1.65in]{press1}
\caption{{\footnotesize{Behaviours of the energy densities $\rho_\gamma(t)$ (left
plot) and $\rho_{\phi}(t)$ (right plot) for different values of
the barotropic parameter $\gamma$. We have considered the same initial
energy densities for the barotropic fluid and the tachyon field. At
the bounce we have three different scenarios concerning the matter
content dominance. When $\gamma\approx1$ the system exhibits a tracking
solution. If $\gamma>1$ the fluid is dominant, whereas for $\gamma<1$
it is the tachyon field that is dominant. 
%In the lower center panel we represent
%the evolution of effective pressure, for the three solutions, until
%the bounce is reached.
}}}

\label{F-Endens} 
\end{figure*}

Figure \ref{F-scalef} show the behaviour of the scale factor. Therein
we observe that in the limit $\rho\rightarrow\rho_{\text{crit}}$, when
the Hubble rate vanishes, the classical singularity is replaced by
a bounce (cf. figure \ref{F-scalef}). In figure \ref{F-Endens} we
represent the energy densities, $\rho_\gamma(t)$ (left plot) and $\rho_{\phi}(t)$
(center plot), for different values of the barotropic parameter $\gamma$
at the bounce. We see that three scenarios have to be considered.
When the energy densities of the tachyon and of the fluid scale exactly
at the same power of the scale factor, namely 
\begin{equation}
\rho_{\phi}\ \approx\ \rho_{\gamma}\ \approx\ \rho_{0}\, a^{-3\gamma},\label{trackb-1}
\end{equation}
then the semiclassical solutions display a {\em tracking} behaviour \cite{RLazkoz,Liddle}.
Numerical analysis shows that this happens when the barotropic parameter
is approximately $\gamma\sim1$, that is, the collapse matter content
acts like dust. From Eq. (\ref{trackb-1}), we have $a_{\textrm{crit}}=\left[\rho_{\textrm{crit}}/\left(2\rho_{0}\right)\right]^{-1/3}$
at the bounce for the tracking solution. In the case where $\gamma>1$
the solution at the bounce is fluid dominated, whereas for $\gamma<1$,
the tachyon field is the dominant component of the energy density
content of the system; in the limit case where the barotropic 
fluid is absent (i.e. $\gamma =0$ and $\rho_\gamma=0$) the collapsing system is {\em purely tachyonic}.
This solution will further enhance the characteristics of the  tachyon dominated solutions.

From figure \ref{F-Endens} we also observe that, starting
from very low values of the energy density (classical regime), a system
that is {\em fluid dominated} reaches the bounce faster than a system that
is {\em tachyon dominated}. This seems to point to the fact that a fluid
dominant solution will drive the energy density until its critical
value more efficiently than when the tachyon field is dominant. 
In the absence of the barotropic fluid, 
the matter content of the collapse is purely tachyonic, so
the tachyon  density $\rho_{\phi}$ will reach the critical value $\rho_{\rm crit}$ at the bounce.
In this case, since there exist no  barotropic fluid to push the collapse, the time required to reach the bounce will be maximized.
In order to explain this result, let us consider what happens to the
total pressure $p_{\phi}+p_{\gamma}$ for each solutions discussed
in this section. When the tachyon field is dominant, for $\gamma<1$,
the total pressure 
\begin{equation}
p\ =\ -V(\phi)\sqrt{1-\dot{\phi}^{2}}+\left(\gamma-1\right)\rho_{\gamma}\ ,
\label{total press.}
\end{equation}
is negative until the collapsing body reaches the bounce. In fact,
near the bounce $\dot{\phi}\rightarrow1$ and eq. (\ref{total press.})
becomes $p\thickapprox\left(\gamma-1\right)\rho_{\gamma} <0$.
For the tracking solution $\left(\gamma\sim1\right)$, we have $p\sim0$
and the matter content behaves as dust. Finally, for the fluid dominated
solutions, the total pressure is approximately $p\thickapprox\left(\gamma-1\right)\rho_{\gamma}$,
which is positive because in this case $\gamma>1$. Consequently,
in this last scenario, the positive pressure drives the fluid dominant
content of the energy density rapidly towards its critical value, $\rho_{\text{crit}}$,
at the semiclassical bounce.

\begin{figure*}[t]
\centering\includegraphics[height=2.2in]{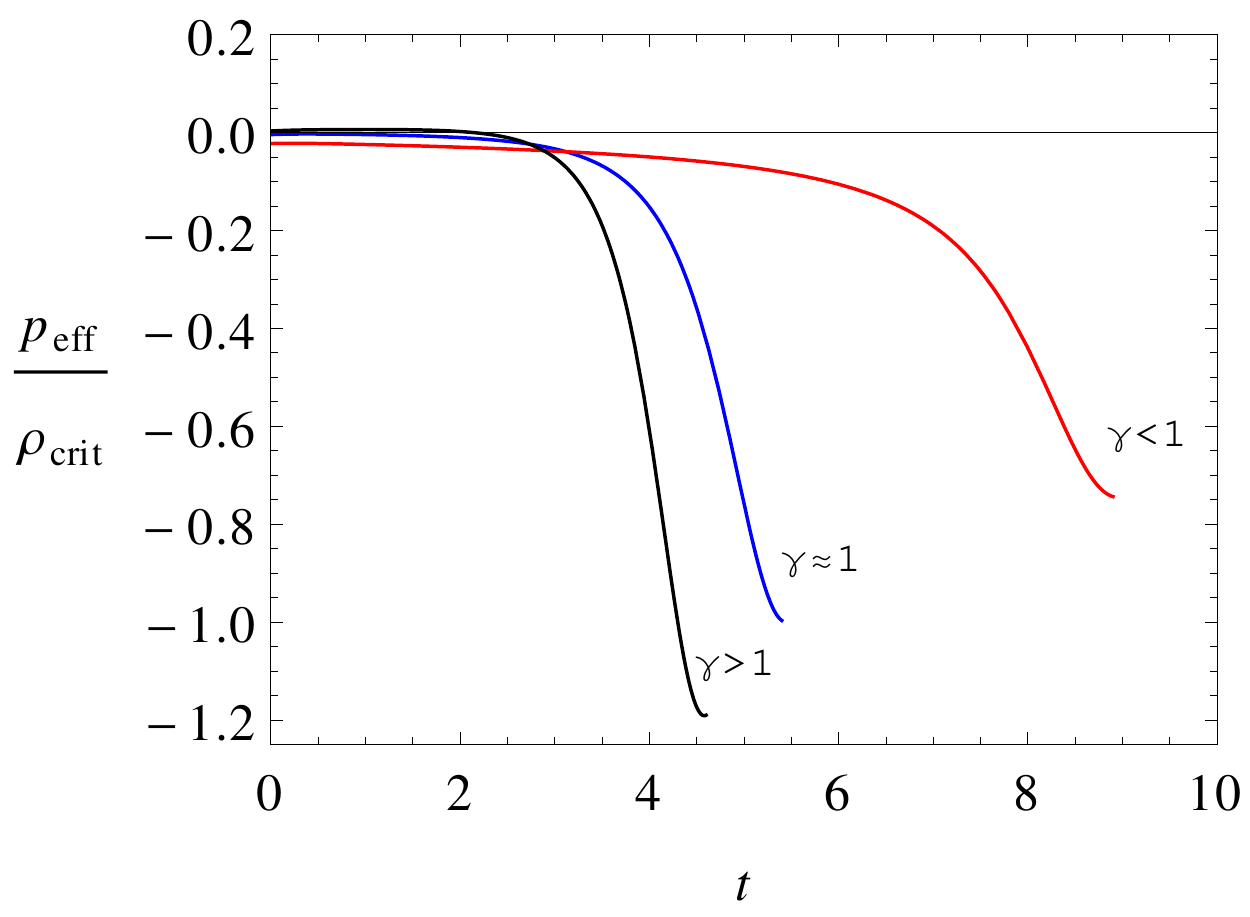}
\caption{{\footnotesize{In this plot we represent
the evolution of the effective pressure $p_{\rm eff}$, for the three cases $\gamma>1$, $\gamma\approx1$  and $\gamma<1$  until
the bounce is reached.}}}

\label{F-Endens-2} 
\end{figure*}

In addition, when we consider Eq. (\ref{Peff}) for the effective
pressure, in particular its value at the bounce (where $\rho\rightarrow\rho_{\textrm{crit}}$),
\begin{equation}
\begin{cases}
p_{\textrm{eff}}^{\phi}\ \approx\ -\left(\gamma-1\right)\rho_{\gamma}-\rho_{\textrm{crit}}\ , & \quad \quad\gamma<1\\
p_{\textrm{eff}}^{tr}\ \approx\ -\rho_{\textrm{crit}}\ ,  & \quad \quad\gamma\approx1\\
p_{\textrm{eff}}^{\gamma}\ \approx\ -\left(\gamma-1\right)\rho_{\gamma}-\rho_{\textrm{crit}}\ ,  &\quad \quad\gamma>1
\end{cases}
\end{equation}
we can establish that $p_{\textrm{eff}}^{\gamma}<p_{\textrm{eff}}^{tr}<p_{\textrm{eff}}^{\phi}<0$
(see  figure \ref{F-Endens-2}). In this plot we have that
for the fluid dominated solution, the effective pressure start at
a positive value (pushing the density of collapsing system to increase rapidly
towards its critical value $\rho\rightarrow\rho_{\text{crit}}$).
However, near the bounce, the effective pressure rapidly switches
to negative values. In contrast, for the tachyon dominated solution,
the effective pressure starts from negative values from the beginning;
this is related to the fact that the initial energy densities of both
the tachyon and barotropic fluid are approximate. Moreover, the change
near the bounce is less pronounced in this last case. Therefore the
evolution of the collapse is slower and the bounce is delayed when
compared to the fluid dominated scenario. It is straightforward to
verify that the tracking solution provides an intermediate context
between the fluid and tachyon dominated solutions.

\subsection{Horizon formation}

\begin{figure*}[t]
\centering\includegraphics[height=1.7in]{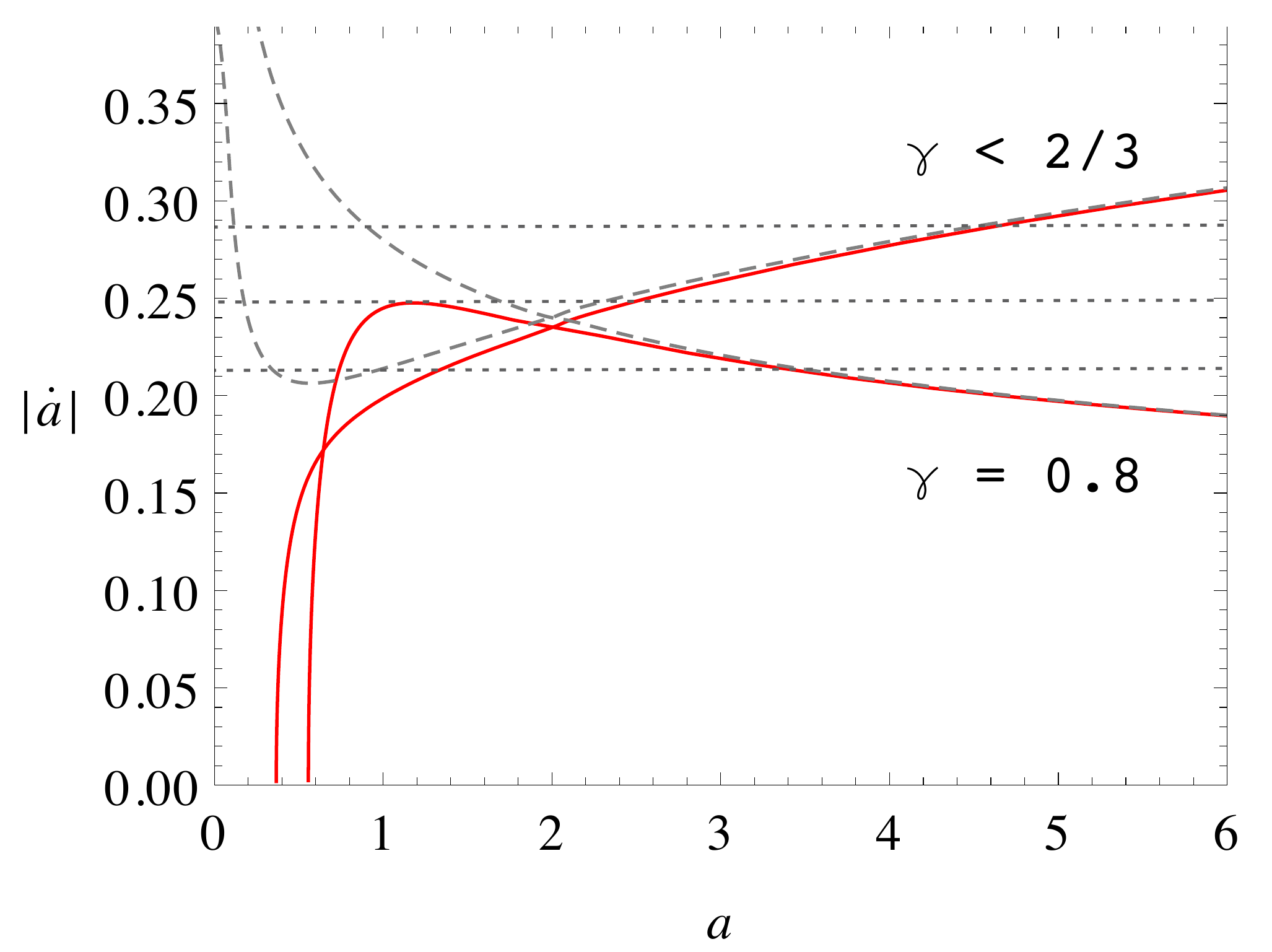}\enskip{}\includegraphics[height=1.7in]{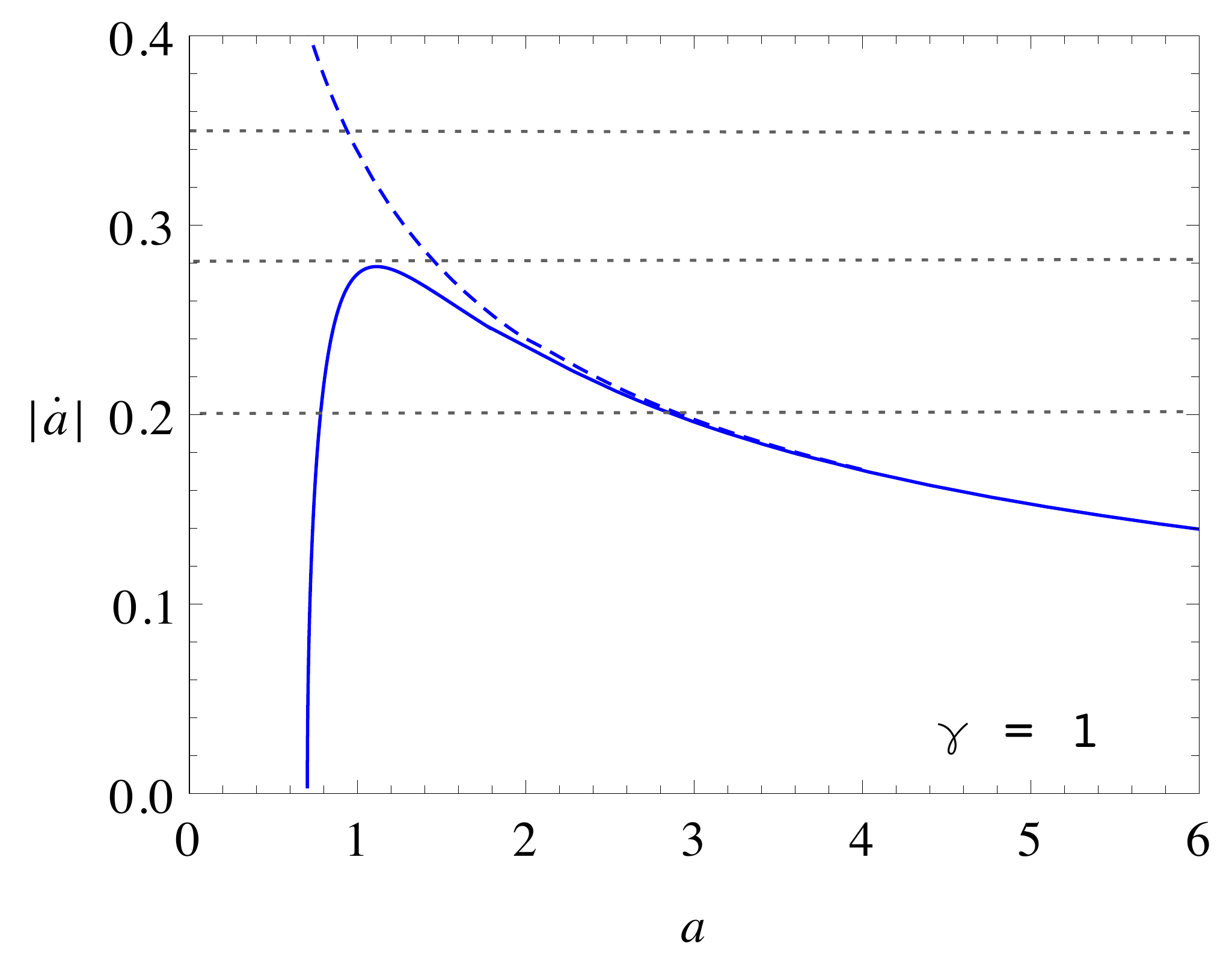}\enskip{}\includegraphics[height=1.7in]{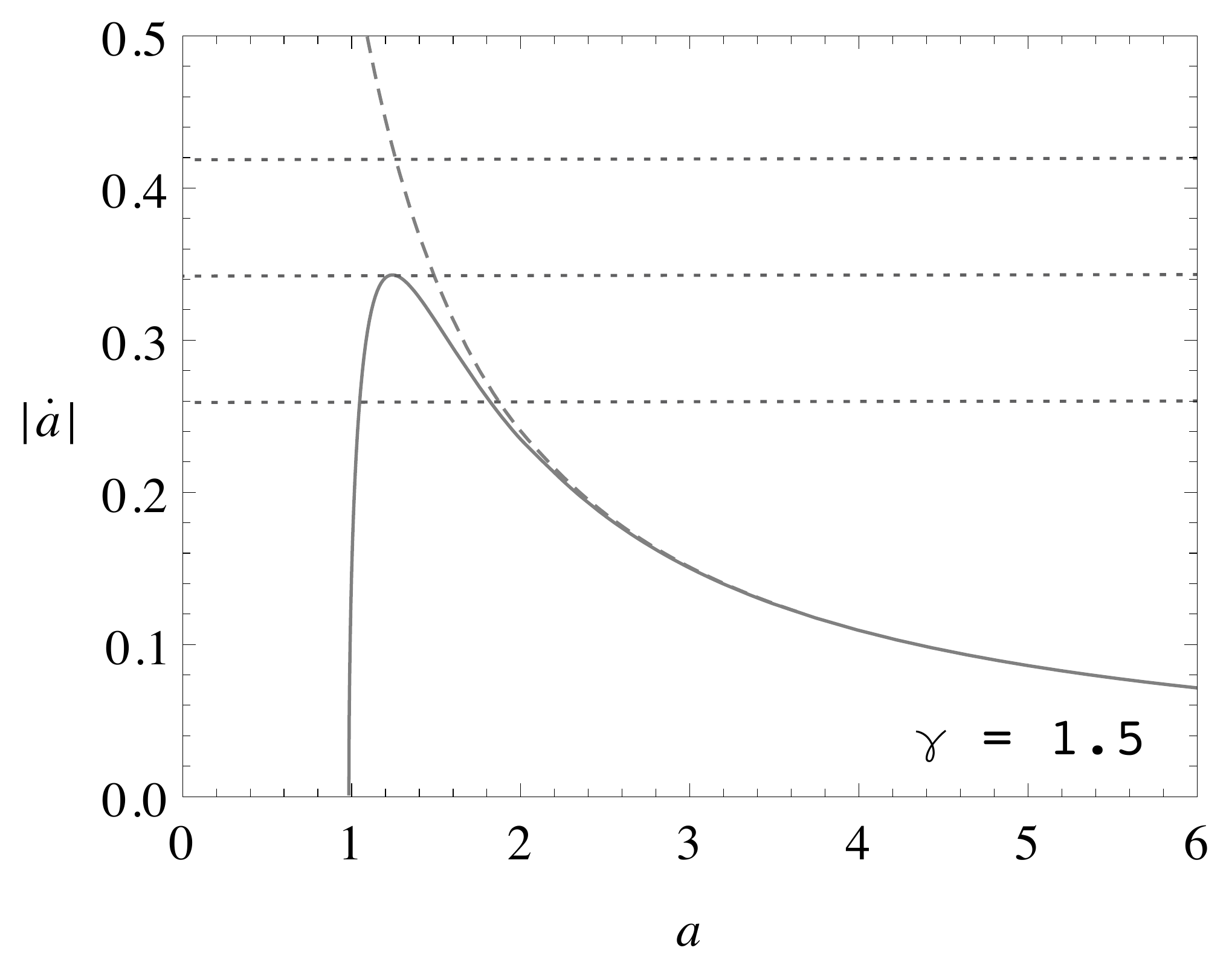}
\caption{{\footnotesize{{The speed of collapse, $|\dot{a}|$, with respect
to the scale factor $a$, in the semiclassical (full line), and classical
(dashed curve) regime. We considered the initial conditions as: $\rho_{0}=0.04$,
$V_{0}=0.001$, $a(0)=2$, $\dot{\phi}\left(0\right)=0.5$ and $\phi_{0}=0.6$.
Top left panel is for a tachyon dominated solution with $\gamma=0.8$ and
$\gamma<2/3$. The top right panel is for a tracking solution with $\gamma=1$.
Finally, the lower center panel is for a fluid dominated solution with $\gamma=1.5$
The dotted lines are for the different values of $r_{b}<r_{\star}$
(upper line), $r_{b}=r_{\star}$ (inner line) and $r_{b}>r_{\star}$
(lower line).}}}}

\label{F-speed-1} 
\end{figure*}

From the equation $\dot{R}^{2}(t,r_{b})=1$ (where $r_{b}$ is the
radius of the boundary shell) we can determine the speed of the collapse,
$|\dot{a}|_{\mathrm{AH}}$, at which horizons form, i.e., $|\dot{a}|_{\mathrm{AH}}=\frac{1}{r_{b}}$.
When the speed of collapse, $|\dot{a}|$, reaches the value $1/r_{b}$,
then an apparent horizon forms. Thus, if the maximum speed $|\dot{a}|_{\text{max}}$
is lower than the critical speed $|\dot{a}|_{\mathrm{AH}}$, no horizon
can form. More precisely, in order to discuss the dynamics of the
trapped region in the perspective of the effective dynamics scenario,
we consider $|\dot{a}|$ from Eq.~(\ref{Friedmann-eff-1a}) to be
equal to $|\dot{a}|_{\mathrm{AH}}=1/r_{b}$. Solving this new equation
for $\rho$ and $a$ we get scale factors and energy densities at
which the horizon forms. Figure \ref{F-speed-1} represents the speed
of the collapse, $|\dot{a}|$, as a function of the scale factor,
reaching the maximum value $|\dot{a}|_{\textrm{max}}$.

The tachyon field equation (\ref{field-eq-1}) implies that $\phi\equiv\phi\left(a\right)$.
Therefore, from Eqs.~(\ref{ener-1})--(\ref{ener-2}) we can also
establish that the total energy density can be expressed as a function
$\rho\equiv\rho\left(a\right)$. Then, we can rewrite $|\dot{a}|=\frac{1}{r_{b}}$
by setting $X:=\rho/\rho_{{\rm crit}}$ and $a^{2}:=f(X)$ as 
\begin{equation}
f\left(X\right)X\left(1-X\right)-A=0\ ,\label{theta2-q2-X}
\end{equation}
where $A:=3/\left(8\pi G\rho_{{\rm crit}}r_{b}^{2}\right)$ is a constant.
The study of roots of the Eq. (\ref{theta2-q2-X}) enables us to get
the values of energy density at which an apparent horizons form. Considering
more closely Eq.~(\ref{theta2-q2-X}), we need to estimate the behaviour
of the function $f(X)$. In figures \ref{F-scalef} and \ref{F-Endens},
we have that $f(X)$ is minimum when $X$ is maximum. It is also expected
that, since $f(X)$ is a monotonically decreasing function near the
bounce, Eq.~(\ref{theta2-q2-X}) is essentially described as a second
order polynomial. Therefore, depending on the initial conditions,
in particular on the choice of the $r_{b}$, three cases can be evaluated,
which correspond to \emph{no} apparent horizon formation ($A/f\left(X\right)>1/4$),
one and two horizons formation ($A/f\left(X\right)\leq1/4$).

Let us introduce a radius $r_{\star}$, defined by 
\begin{equation}
r_{\star}\ :=\ \frac{1}{|\dot{a}|_{\text{max}}}\ .
\end{equation}
We see that $r_{\star}$ determines a \emph{threshold radius} for
the horizon formation; if $r_{b}<r_{\star}$, then no horizon can
form at any stage of the collapse. The case $r_{b}=r_{\star}$ corresponds
to the formation of a dynamical horizon at the boundary of the two
space-time regions \cite{dynHorizon}. Finally, for the case $r_{b}>r_{\star}$
two horizons will form, one inside and the other outside of the collapsing
matter.

The behaviour of the three possible scenarios (tracking solution,
tachyon and fluid dominated solutions) are also represented in figure
\ref{F-speed-1}.  We note  that only one horizon
forms for some particular tachyon dominated solutions. Therefore,
for these solutions the bounce will be covered by an horizon. In order
to further clarify this aspect, we note that when more than one horizon
forms, the speed of the collapse $\dot{a}$ must have a local maximum.
In that case, the acceleration must be $\ddot{a}=0$, and from equations
(\ref{Friedmann-eff-1a})--(\ref{LFriedmann2}) we can determine that
this local maximum can be found by imposing 
\begin{equation}
\frac{\ddot{a}}{a}\ = \ \dot{H}+H^{2}\ =\ -\frac{\kappa}{6} \left(\rho_{\rm crit} + 3p_{\rm crit} \right)\ =\ 0\ .
%-\dfrac{\rho_{\textrm{eff}}}{6}-\dfrac{p_{\textrm{eff}}}{2}=0\ .
\label{accel-1}
\end{equation}
This last condition, being equivalent to $\rho_{\textrm{eff}}=-3\, p_{\textrm{eff}}$,
must be closely monitored for the three different solutions  discussed
in this section. For the fluid dominated solution, and since the
effective pressure starts from positive values and evolve to negatives
one near the bounce, it is straightforward to verify that the function
$-3\, p_{\textrm{eff}}$ must intersect $\rho_{\textrm{eff}}$ at
some point before reaching the bounce. For the tracking solution we
can use the same argument but with an initial effective pressure starting
near zero and reaching $3\,\rho_{\textrm{crit}}$ at the bounce. Finally,
the case of the tachyon dominated solution depends on the value of
the barotropic parameter $\gamma$. When the initial values for the
effective pressure and energy densities are $p_{\textrm{eff}}^{\phi}\approx-\left(\gamma-1\right)\rho_{\gamma0}$
and $\rho_{\textrm{eff}}^{\phi}\approx\rho_{\gamma0}>\rho_{\phi0}$,
respectively; then, if $\gamma>2/3$, the argument given for the tracking
and fluid dominated solution is also valid for this case. However,
if $\gamma<2/3$, there will be at the most one horizon forming. Besides
taking $\gamma>2/3$, if we consider an unbalanced initial energy
density, with the tachyon being slightly dominant, i.e., $\rho_{\phi0}\geq\rho_{\gamma0}$,
a local maximum for $\dot{a}$ will also be present.

\subsection{Exterior geometry}

\begin{figure*}[t]
\centering\includegraphics[height=2.in]{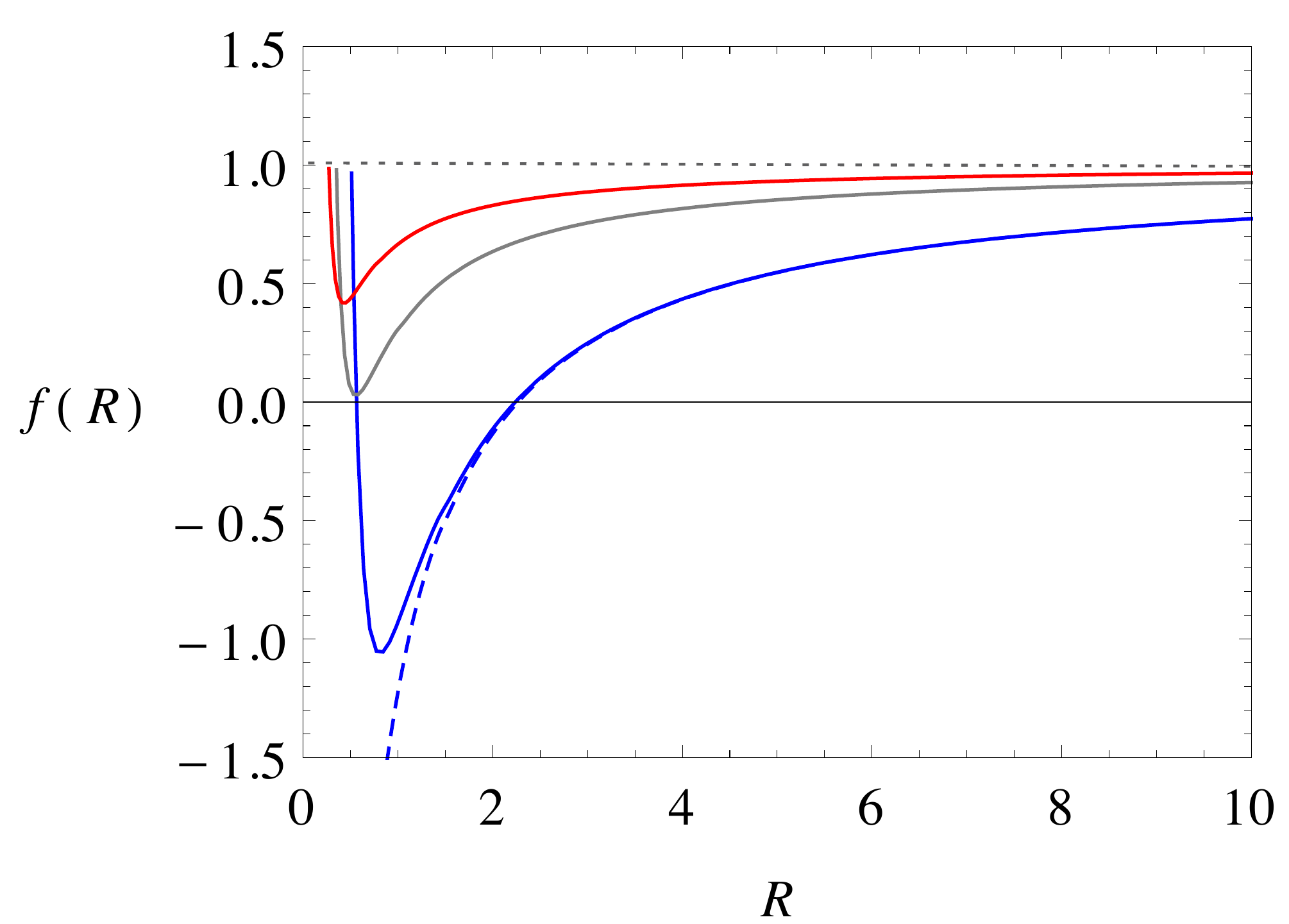}\quad{} \quad \includegraphics[height=2.in]{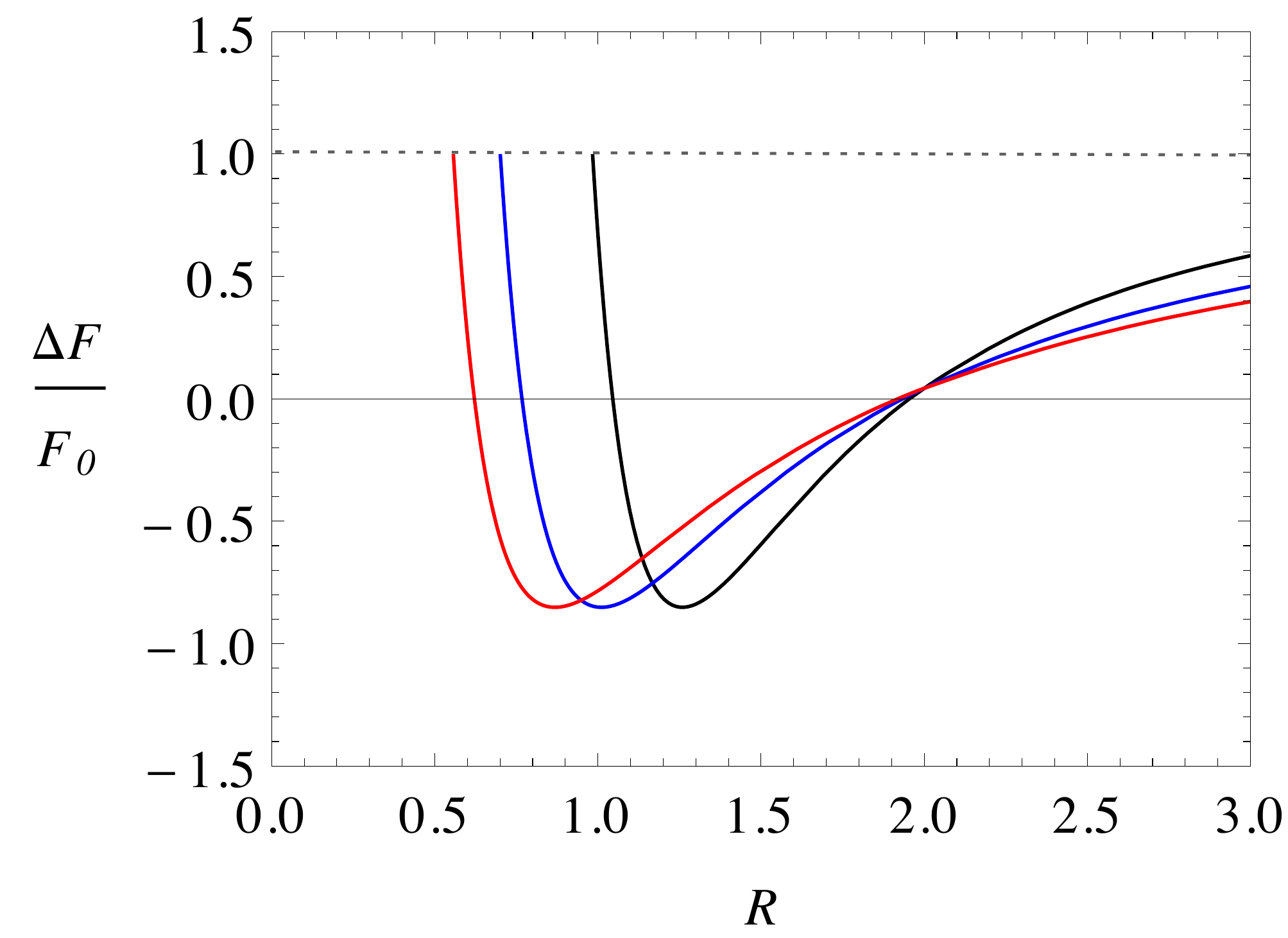}
\caption{{\footnotesize{The left plot represents the tracking solution ($\gamma\approx1$)
boundary function, $f(R)$, with respect to the radius $R$, in the
semiclassical (full line), and classical (dashed curve) regime. We
considered the initial conditions as: $\rho_{0}=0.04$, $V_{0}=0.001$,
$a(0)=2$, $\dot{\phi}\left(0\right)=0.5$ and $\phi_{0}=0.6$. The
right plot shows the behaviour of the mass loss $\Delta F/F$, as
a function of area radius $R$. In this last plot we present the behaviour
of the tachyon dominated solution (reaching $\Delta F/F\rightarrow1$
at the smaller $R$), the tracking solution (reaching $\Delta F/F\rightarrow1$
at the intermediate $R$) and the fluid dominated solution (reaching
$\Delta F/F\rightarrow1$ at the bigger $R$).}}}

\label{Bound-f(R)} 
\end{figure*}

Finally, the discussion of the final outcomes  related
to the semiclassical solutions  follows the one made in Ref. \cite{us:2013c}.
In  this previous work, it is described that the
fate of the collapsing star (with a massless scalar field as matter source) whose shell radius is less than the threshold
radius $r_{\star}$ points to the existence of an energy flux radiated
away from the interior space-time and reaching the distant observer.
In the present study, for a collapsing system whose
initial boundary radius $r_{b}$ is less than $r_{\star}$, we analyze
the resulting mass loss due to the semiclassical modified interior
geometry. In particular, this analysis is only carried for the tracking
solution or fluid dominated scenario, since the tachyon dominated
solution develops no more than one horizon, for $\gamma<2/3$ and $\rho_{\gamma0}\geq\rho_{\phi0}$,
before reaching the bounce. Let us designate the initial mass function
at scales $\rho\ll\rho_{{\rm crit}}$, i.e, in the classical regime,
as $F_{0}=(8\pi G/3)\rho_{0}R_{0}^{3}$, with $\rho_{0}=\rho_{\phi0}+\rho_{\gamma0}$.
For $\rho\lesssim\rho_{{\rm crit}}$ (in the semiclassical regime)
we have, instead, an effective mass function $F_{\text{eff}}$ given
by Eq. (\ref{massF-eff}). Then, the (quantum geometrical) mass loss,
$\Delta F/F_{0}$ (where $\Delta F=F_{0}-F_{\text{eff}}$), for any
shell, is provided by the following expression: 
\begin{align}
\frac{\triangle F}{F(a_{0})}\  =\ 1-\frac{F_{\text{eff}}}{F_{0}} \  =\ 1-\sqrt{\frac{\rho}{\rho_{0}}}\left(1-\frac{\rho}{\rho_{{\rm cr}it}}\right)\ .
\label{mass-loss-eff}
\end{align}
%\begin{align}
%\frac{\triangle F}{F(a_{0})}\  & =\ 1-\frac{F_{\text{eff}}}{F_{0}}\notag\\
% & =\ 1-\sqrt{\frac{\rho}{\rho_{0}}}\left(1-\frac{\rho}{\rho_{{\rm cr}it}}\right)\ .\label{mass-loss-eff}
%\end{align}
As $\rho$ increases the mass loss decreases positively until it vanishes
at a point. Then, $\Delta F/F$ continues decreasing (negatively)
until it reaches to a minimum at $\rho=\rho_{{\rm crit}}/3$. Henceforth,
in the energy interval $\rho_{{\rm crit}}/3<\rho<\rho_{{\rm crit}}$,
the mass loss increases until the bouncing point at $\rho\rightarrow\rho_{\text{crit}}$,
where $\Delta F/F\rightarrow1$; this means that the quantum gravity
corrections, applied to the interior region, give rise to an outward
flux of energy near the bounce in the semiclassical regime. The previously
described behaviour for the mass loss will be qualitatively identical
with respect to the solution considered. Therefore the tachyon (when
$\gamma>2/3$ or the initial energy densities are $\rho_{\phi0}>\rho_{\gamma0}$),
fluid dominated or tracking solutions will exhibit the same profile
for the mass loss. The only difference between these three cases,
shown in the right plot of figure \ref{Bound-f(R)}, is the value
of the radius where the mass loss reaches the maximum $\Delta F/F\rightarrow1$.
In the last section we discussed the fact that the bounce occurring
in the tachyon dominated solution is delayed compared with the other
solutions. Consequently, the bounce (where $\Delta F/F\rightarrow1$)
take place for a smaller value of the radius $R$.

In the other case, where $r_{b}\geq r_{\star}$, in which one or two
horizon form, the exterior geometry can be obtained by matching the
interior to a generalized Vaidya exterior geometry at the boundary
$r_{b}$ of the cloud. Following the method provided in Ref.~\cite{us:2013a},
we can write the exterior metric in advanced null coordinates $(v,R)$
as 
\begin{equation}
ds_{{\rm (ext)}}^{2}\ =\ -f(R,v)dv^{2}-2dvdR+R^{2}d\Omega^{2}\ ,
\end{equation}
where the exterior function is given by $f(v,R)=1-2Gm(v,R)/R$. By
applying the matching conditions at the boundary $r_{b}$ we have
that 
\begin{equation}
m(v,R)\ =\ M-\frac{3}{4\pi\rho_{{\rm crit}}}\frac{M^{2}}{R^{3}}
\end{equation}
where we have defined $M:=(4\pi/3)\rho R^{3}$ is the mass within
the volume $R^{3}$.  For a fluid dominated solution  we
have 
\begin{equation}
M\ \approx \ \frac{4\pi}{3}\rho_{0}R_{0}^{3\gamma}R^{-3(\gamma-1)}\ .
\label{mass-2}
\end{equation}
In the limit case $\gamma\sim1$, i.e. the tracking solution, Eq.~(\ref{mass-2}) reduces to
\begin{align}
M\ =\ M_{0}\ =\ \left(4\pi/3\right)\rho_{0}R_{0}^{3}\ ,
\end{align}
which presents a modified Oppenheimer-Snyder collapse \cite{Oppenheimer} of homogeneous dust matter.
Figure \ref{Bound-f(R)}
shows the numerical behaviour of the boundary function $f(R)$ in
the classical (dashed curve) and semiclassical regime (solid curves)
for the cases of the initial radius $r_{b}<r_{\star}$ and $r_{b}\geq r_{\star}$.
The later shows the behaviour of an exotic non-singular black hole
geometry which is different from  its classical counterpart.
Similarly, for a tachyon dominated case with $\gamma<1$ (or a purely tachyonic collapse), the mass $M\approx \left(4\pi/3\right)\rho_{\phi}R^{3}$ reads
\begin{align}
M(\phi)\ \approx \  \frac{4\pi R^3}{3}\frac{V(\phi)}{\sqrt{1-\dot{\phi}^2}} \ .
\end{align}
It should be noticed that the energy density $\rho_{\phi}$ is upper bounded by the critical density $\rho_{\rm crit}$, thus, the mass $M$ is always finite.

\section{Conclusion and discussion}

\label{conclusion}

In this paper we employed an effective scenario imported from LQG,
namely the ``holonomy'' corrections to the dynamics of the gravitational
collapse whose matter content involves a self interacting tachyon
field and a barotropic fluid. Our aim was to enlarge the discussion
on tachyon field gravitational collapse. More concretely, 
on the one hand, extending the scope analyzed
in Ref. \cite{us:2013a}, by investigating how the quantum gravity
correction term $-\rho^{2}/\rho_{{\rm crit}}$, can alter the fate
of the collapse. Using a dynamical system analysis, we subsequently
found a class of solutions. 
Our semiclassical analysis showed that, the corresponding
stable fixed point (attractor) solutions in the classical general
relativistic collapse become saddle points in our semiclassical
collapse; hence, the classical black hole and naked singularities
produced  in Ref.~\cite{us:2013a} are no longer present within the 
loop semiclassical regime studied in this paper. 
We found conditions to define a tachyon
or fluid dominated regimes close to the bounce, depending on the value of
the barotropic parameter of the fluid. The transition from one regime
to the other shows the emergence of a tracking solution where the
collapsing matter behaves as dust. 
It was also observed that, in
scenarios  with similar  initial conditions, a fluid dominated
solution  drives  the energy density of the collapsing cloud until the critical value $\rho_{\rm crit}$
more rapidly than a tachyon dominated  solution. 

Our analysis  provides elements to contrast  the correction features of holonomy   to those given by  inverse triad modification  \cite{us:2013b}, 
as far as tachyon scalar field  gravitational collapse is concerned.
Nevertheless, the issue of inverse triad effects is quite subtle in LQC; it has been shown that for space-times with non-compact topology they can not be consistently incorporated. 
%For compact topologies they can be consistently incorporated. 
(We suggest the reader to see Refs.~\citep{LQC-AS,Rev1,Rev2} where further technical details  on the problem  with inverse scale factor effects is discussed.) 
In addition, it is important to mention that for space-times with spatial curvature there is no problem in including inverse scale factor effects (see Ref.  \cite{Rev3}).
In the aforementioned case of a gravitational collapse modified by inverse
triad corrections, the energy density of collapsing matter source
is a decreasing function of time towards the star center and remains
constant there \cite{us:2013b}. 
However, in the presence of the holonomy
corrections, the quadratic density modifications provide an upper
limit $\rho_{{\rm crit}}$ for the energy density $\rho$ of the collapsing
matter, indicating that  the herein semiclassical collapse leads
to a non-singular bounce at the critical density $\rho=\rho_{{\rm crit}}$.
The mass loss obtained by the inverse triad scenario \cite{us:2013b},
was characterised by the reduction of energy density and mass function
towards the star center, which leads to an outward energy flux. However,
in the present semiclassical model, for a particular range of the
radius of the boundary shell (i.e., $r_{b}<r_{\star}$), the energy
density and the mass function growth of the collapsing cloud is followed by the effective 
mass loss reduction near the bounce, which subsequently gives rise to an outward energy
flux at the collapse end state. 
In addition, a detailed dynamical
system analysis in the inverse triad scenario predicted only one stable
solution which allowed a barotropic fluid with a parameter satisfying
the range $8/5\leq\gamma<2$ (where the superior limit is a consequence
of the energy conditions imposed in the classical setting, discussed
in Ref. \cite{us:2013a}). In contrast to the case studied in Ref. \cite{us:2013b}, the
barotropic fluid parameter is less constrained and solutions with
$0<\gamma<2$, as predicted by the classical model, are allowed to evolve until the bounce.

We further investigated, by means of numerical studies, the evolution
of trapped surfaces during the collapse in order to determine its
final state. We found a threshold radius for the collapsing matter
cloud in order to form a black hole at late time stages. The physical
modifications related to the semiclassical regime provided three cases
for the trapped surfaces formation, depending on the initial conditions
of the collapsing star. In particular, our solutions showed that,
if the initial boundary radius of the collapsing cloud is less than
a threshold radius, namely $r_{\star}$, \emph{no} horizon forms during
the collapse, whereas for the radius equal and larger than the $r_{\star}$,
one and two horizons form, respectively. 
It is worthy to mention that,
for the tachyon dominated solutions,
the previous scenario only happens when the barotropic parameter is $\gamma>2/3$
or the initial energy densities satisfy $\rho_{\phi0}\geq\rho_{\gamma0}$.
When  $\gamma<2/3$ and $\rho_{\phi0}<\rho_{\gamma0}$, not more than one apparent horizon
forms.

Our effective scenario, in the presence of a tachyon field joined to a
barotropic fluid, share a few common features to the one where, instead, a homogeneous
massless scalar field was considered for the collapsing matter content
\cite{us:2013c}. In both contexts, and in the particular case in which no horizon
forms, it is shown that, as the collapse evolves, the energy density
increases towards a maximum value $\rho_{\textrm{crit}}$ at the bounce.
Moreover, in these semiclassical scenarios, the effective energy density
reduction leads to a positive mass loss near the bounce. This results in
a positive luminosity near the bounce which  gives rise to
an outward energy flux from the interior region, that  may reach to
a distant observer. 
In addition, in the cases in which one or two
horizons form, the resulting exterior geometry corresponds to exotic
non-singular black holes which are different from the Schwarzschild
one (see also Refs.~\citep{Bojowald:2005,VHussain} for such a black
hole formation).

In our collapsing system, additional physical
situations  might arise from the interplay between the tachyon and the barotropic fluid. 
Namely, the distinct matter dominance regimes yield a wider variety of outcomes, 
either for the horizon formation, 
or when the efficiency to reach the semiclassical bounce is concerned.
Including an  interaction 
term $\Gamma_{{\rm int}}$, that could account for a transfer of energy
between the tachyon field and the barotropic fluid, is expected to change the
results presented here. In fact, it is reasonable to anticipate that the conditions
required to define the fluid or tachyon dominated regimes will certainly be less simple.
The emergence of these regimes will certainly reflect more than just the variation 
of the barotropic parameter. Another interesting question to be addressed, in this  context,
is related to the existence of a tracking behaviour and under which conditions it becomes possible.
In fact, from the results presented here, the emergence of a tracking behaviour
seems to occur at the transition between the fluid and the tachyon dominated regimes. Would this situation
be maintained in the interacting case?
The answers to these questions will also have an impact on the discussion
about the horizon formation, since it is closely related to the matter
dominance present at the bounce. Therefore, one interesting extension to this work
will be to consider an interaction between the tachyon and the barotropic fluid.

The qualitative picture depicted from our toy model is strongly dependent
on the choice of a  homogeneous interior space-time. Nevertheless,
in a realistic collapsing scenario one should employ a more general
inhomogeneous setting (see Refs. \citep{Bojowald:2006,Campiglia:2007},
where a detailed introduction to recent techniques to handle inhomogeneous
systems provides the ingredients on how to extend the limited homogeneous
case). 
When we apply homogeneous techniques, the quantum effects
are restricted to the interior space-time, whereas,  the outer  space-time region
is assumed to be a generalised Vaidya metric defined by classical
general relativity. Some imprint of the interior quantum effects are
transported to the outside, by imposing suitable matching conditions
at the boundary surface, where it enters the Vaidya solution effectively
through a nonstandard energy-momentum tensor. This procedure is also
restricted by the fact that  a full inhomogeneous quantization,
also covering the exterior region, is expected to provide significant
modifications to the space-time structure. However, some indications
on how the matter  content might  affect the bounce
scenario may still be valid in a more general inhomogeneous setting.

\section{Acknowledgments}

Y.T.  was supported by FCT (Portugal) through the fellowship SFRH/BD/43709/2008 and a grant from CNPq (Brazil).
This research work was supported by the grants CERN/FP/123609/2011
and CERN/FP/123618/2011 and PEst-OE/MAT/UI0212/2014. 

%%%%%%%%%%%%%%%%%%%%%%%%%%%%%%%%%%%%%%%


\begin{thebibliography}{10}
\bibitem{Joshi:2007} P. Joshi, \emph{Gravitational Collapse and Space-Time
Singularities}, (Cambridge University Press, England, 2007).

\bibitem{Penrose:1965} R. Penrose, Phys. Rev. Lett. \textbf{14},
57 (1965); S. W. Hawking, Proc. R. Soc. A \textbf{300}, 187 (1967);
S. W. Hawking and R. Penrose, Proc. R. Soc. A \textbf{314}, 529 (1970).

\bibitem{Hawking:1974} S. W. Hawking and G. F. R. Ellis, \emph{The
Large Scale Structure of Space-Time}, (Cambridge University Press,
England 1974).

\bibitem{Ashtekar:2004} A. Ashtekar and J. Lewandowski, Class. Quantum
Grav. \textbf{21}, R53 (2004); T. Thiemann, \emph{Introduction to
Modern Canonical Quantum General Relativity} (Cambridge University
Press, Cambridge, England, 2007).

\bibitem{Ashtekar:2005} A. Ashtekar, M. Bojowald and J. Lewandowski,
Adv. Theor. Math. Phys. \textbf{7}, 233 (2003).

\bibitem{LQC-AS} A. Ashtekar and P. Singh, \emph{Loop quantum cosmology:
a status report}, Class. Quantum Grav. \textbf{28}, 213001 (2011).

\bibitem{Bojowald:2010} M. Bojowald, \emph{Canonical Gravity and
Applications: Cosmology, Black Holes, and Quantum Gravity} (Cambridge
University Press, Cambridge, England, 2010).

\bibitem{Bojowald:2005} M. Bojowald, R. Goswami, R. Maartens, and
P. Singh, Phys. Rev. Lett. \textbf{95}, 091302 (2005).

\bibitem{Goswami:2006} R. Goswami, P. S. Joshi and P. Singh, Phys.
Rev. Lett. \textbf{96}, 031302 (2006).

\bibitem{Ashtekar:2006} A. Ashtekar, T. Pawlowski and P. Singh, Phys.
Rev. Lett. \textbf{96}, 141301 (2006).

\bibitem{us:2013c} Y. Tavakoli, J. Marto and A. Dapor, Int. J. Mod.
Phys. D \textbf{27}, No. 7, 1450061 (2014).

\bibitem{Modesto:2013} C. Bambi, D. Malafarina and L. Modesto, Phys.
Rev. D \textbf{88}, 044009 (2013).

\bibitem{us:2013a} Y. Tavakoli, J. Marto, A. Hadi Ziaie and P. Vargas
Moniz, Gen. Rel. Grav. \textbf{45}, 819 (2013).

\bibitem{Li:2009} L.-F. Li and J.-Y. Zhu, Phys. Rev. D \textbf{79},
124011 (2009).

\bibitem{us:2013b} Y. Tavakoli, J. Marto, A. H. Ziaie and P. V. Moniz,
Phys. Rev. D \textbf{87}, 024042 (2013).

%\bibitem{PSingh:2006} P. Singh, Phys. Rev. D \textbf{73}, 063508
%(2006).

\bibitem{Taveras:2006} V. Taveras, IGPG preprint (2006).

\bibitem{TB-Int1} R. Herrera, D. Pav\'on and W. Zimdahl, Gen. Rel. Grav. \textbf{36}, 9 (2004).

\bibitem{TB-Int2} W. Chakraborty and U. Debnath, Astrophys. and Space Science
\textbf{315}, 167  (2008).

\bibitem{TB-Int3} M. R. Setare, J. Sadeghi and A. R. Amani, Phys. Lett. B \textbf{673}, 4 (2009).

%\bibitem{TB-Int4} H. Amirhashchi, A. Pradhan and B. Saha, Chinese Phys. Lett. \textbf{28}, 3 (2011).

%\bibitem{TB-Int5} B. Saha,  H. Amirhashchi and A. Pradhan,  Astrophysics and Space Science \textbf{342}, 1 (2012).

\bibitem{NSDyn} H. K. Khalil, \emph{Nonlinear Systems}, 2nd edition
(Englewood Cliffs. NJ: Prentice Hall, 1996) pp 167-177.

\bibitem{RLazkoz} Juan M. Aguirregabiria and Ruth Lazkoz, Phys. Rev.
D \textbf{69}, 123502 (2004).

\bibitem{Liddle} A. R. Liddle and R. J. Scherrer, Phys. Rev. D \textbf{59},
023509 (1998).

%\bibitem{MS} S. A. Hayward, Phys. Rev. D \textbf{53}, 1938 (1996).

\bibitem{dynHorizon} S. A. Hayward, Phys. Rev. D \textbf{49}, 6467
(1994); A. Ashtekar and B. Krishnan, Phys. Rev. Lett. \textbf{89},
261101 (2002).

\bibitem{VHussain} Benjamin K. Tippett and Viqar Husain, Phys. Rev.
D \textbf{84}, 104031 (2011).

\bibitem{Bojowald:2006} M. Bojowald, R. Swiderski, Class. Quantum
Grav. \textbf{23}, 2129 (2006).

\bibitem{Campiglia:2007} M. Campiglia, R. Gambini and J. Pullin,
Class. Quantum Grav. \textbf{24}, 3649 (2007).

% \bibitem{Bojowald:2012} M. Bojowald, \emph{Quantum cosmology: effective
% theory}, Class. Quantum Grav. \textbf{29}, 213001 (2012).

% \bibitem{Bojowald:2011} M. Bojowald, G. M. Paily, J. D. Reyes and
% R. Tibrewala, Class. Quantum Grav. \textbf{28}, 185006 (2011).

\bibitem{Oppenheimer}  J. R. Oppenheimer and H. Snyder, Phys. Rev. {\bf 56}, 455 (1939).

\bibitem{Rev1} A. Ashtekar, T. Pawlowski and P. Singh, Phys. Rev. D \textbf{74}, 084003 (2006).

\bibitem{Rev2} A. Ashtekar, T. Pawlowski, P. Singh and K. Vandersloot, Phys. Rev. D \textbf{75}, 024035 (2007).

\bibitem{Rev3} B. Gupt and P. Singh, Phys. Rev. D \textbf{85}, 044011 (2012).



\end{thebibliography}
\end{document}